\documentclass[twocolumn]{aastex62}

\usepackage{graphicx}
\graphicspath{{./Figures/}}
\usepackage{subfigmat}
\usepackage{multirow}

\received{-}
\revised{-}
\accepted{-}\shorttitle{Radiative Transfer modeling of EC 53}
\shortauthors{G. Baek et al.}
\begin{document}

\title{Radiative Transfer modeling of EC 53: An Episodically Accreting Class I Young Stellar Object}
\author{Giseon Baek}
\affiliation{School of Space Research and Institute of Natural Sciences, Kyung Hee University, 1732 Deogyeong-daero, Giheung-gu, Yongin-si, Gyeonggi-do 446-701, Korea;  jeongeun.lee@khu.ac.kr, giseon@khu.ac.kr}

\author{Benjamin A. MacFarlane}
\affiliation{Jeremiah Horrocks Institute for Mathematics, Physics and Astronomy, University of Central Lancashire, Preston, PR1 2HE, UK}

\author{Jeong-Eun Lee}
\affiliation{School of Space Research and Institute of Natural Sciences, Kyung Hee University, 1732 Deogyeong-daero, Giheung-gu, Yongin-si, Gyeonggi-do 446-701, Korea;  jeongeun.lee@khu.ac.kr, giseon@khu.ac.kr}

\author{Dimitris Stamatellos}
\affiliation{Jeremiah Horrocks Institute for Mathematics, Physics and Astronomy, University of Central Lancashire, Preston, PR1 2HE, UK}

\author{Gregory Herczeg}
\affiliation{Kavli Institute for Astronomy and Astrophysics, Peking University, Yiheyuan 5, Haidian Qu, 100871 Beijing, China}

\author{Doug Johnstone}
\affiliation{NRC Herzberg Astronomy and Astrophysics, 5071 West Saanich Rd, Victoria, BC, V9E 2E7, Canada }
\affiliation{Department of Physics and Astronomy, University of Victoria, Victoria, BC, V8P 1A1, Canada}

\author{Carlos Contreras Pe\~{n}a}
\affiliation{School of Physics, Astrophysics Group, University of Exeter, Stocker Road, Exeter EX4 4QL, UK}

\author{Watson Varricatt}
\affiliation{Institute for Astronomy, University of Hawaii, 640 N. A\'{o}hoku Place, Hilo, HI 96720, USA}

\author{Klaus W. Hodapp}
\affiliation{Institute for Astronomy, University of Hawaii, 640 N. A\'{o}hoku Place, Hilo, HI 96720, USA}

\author{Huei-Ru Vivien Chen}
\affiliation{Department of Physics and Institute of Astronomy, National Tsing Hua University, Taiwan}

\author{Sung-Ju Kang}
\affiliation{ Korea Astronomy and Space Science Institute, 776 Daedeokdae-ro, Yuseong-gu, Daejeon 34055, Republic of Korea}

\begin{abstract}
In the episodic accretion scenario, a large fraction of the protostellar mass accretes during repeated and large bursts of accretion. Since outbursts on protostars are typically identified at specific wavelengths, interpreting these outbursts requires converting this change in flux to a change in total luminosity. The Class I young stellar object EC 53 in the Serpens Main cloud has undergone repeated increases in brightness at 850 $\micron$ that are likely caused by bursts of accretion. 
In this study, we perform two- and three-dimensional continuum radiative transfer modeling to quantify the internal luminosity rise in EC 53 that corresponds to the factor of $\sim$1.5 enhancement in flux at 850 $\micron$. We model the spectral energy distribution and radial intensity profile in both the quiescent and outburst phases. The internal luminosity in the outburst phase is $\sim 3.3$ times brighter than the luminosity in the quiescent phase. 
The radial intensity profile analysis demonstrates that the detected sub-mm flux variation of EC 53 comes from the heated envelope by the accretion burst. 
We also find that the role of external heating of the EC 53 envelope by the interstellar radiation field is insignificant. \\
\end{abstract}

\keywords{stars: protostars -- stars: variables: general -- accretion -- radiative transfer}

\section{Introduction} \label{sec:introduction}
Historical models of low mass star formation suggest that a protostar grows in mass at a constant rate \citep{shu77,shu87}. 
In these models, at the end of the Class 0 phase ($\sim$ 0.1 Myr), a protostar of 0.5 M$_{\sun}$ should have an accretion rate of $\sim5\times 10^{-6}$  M$_{\sun}\,{\rm yr}^{-1}$, resulting in the accretion luminosity of $\sim$ 25 L$_{\sun}$.
However, \citet{kenyon90} found that  the mean bolometric luminosity of embedded protostars is $\sim 1 \ \text{L}_\odot$. This discrepancy is known as the {\it luminosity problem}. 
The disparity between theoretical predictions and observations was later confirmed with larger sample sizes and with more accurate luminosities from SEDs that include the far-infrared (FIR)\citep{evans09,enoch09,dunham15,fischer19}.

\citet{kenyon90} suggested episodic accretion, which has quiescent accretion phases interspersed with burst accretion phases, as a solution to the luminosity problem.
In this paradigm, the mass of a protostar grows mostly through brief accretion bursts.
The largest outbursts on optically-visible young stellar objects, FU Orionis-type objects (FUors) are characterized by a brightening of $\sim5 \ \text{mag}$ that may last for decades or even centuries \citep{herbig66, herbig77,kenyon00}. The bolometric luminosities of bonafide FUors range from 20--1000\ $\text{L}_\odot$ \citep{connelley18}.

FUor outbursts could be driven by the disk instabilities, including: (i) the thermal instability \citep{bellin95, clarke96, lodato04}, (ii) the gravitational instability \citep{vorobyovbasu05,vorobyovbasu06,vorobyovbasu10a}, and (iii)  a combination of disk gravitational instabilities in the outer disk and the magneto-rotational instability operating in the inner disk \citep{armitage01, zhu09a, zhu09b, zhu10a, zhu10b, Stamatellos:2011a,  martin12, bae14, Mercer:2017a}. External triggers have been also proposed for the driving mechanisms such as binary-disk interactions \citep{reipurth04} or stellar fly-bys \citep{pfalzner08, pfalzner+08}.
Smaller and shorter eruptions also occur and are typically classified as EX Lupus-type objects (EXors), which typically have preburst luminosities of $1-2 \ \text{L}_\odot$, with outburst luminosities rising to tens of $\text{L}_\odot$ \citep{lorenzetti06,audard10,sipos09,aspin10} with durations of weeks to months \citep{coffey04,audard10}.   Such short outbursts are likely caused by the buildup of gas near the star, followed by rapid accretion of the gas onto the star \citep{dangelo10,dangelo12,armitage16}. The classification of eruptions into FUor or EXor outbursts is often unclear \citep[e.g.][]{contreraspena17}.

On the assumption that an FUor outburst is due to the gravitational instabilities in the disk, FUor outbursts may be more frequent and likely more intense during the embedded protostellar phase \citep[e.g.][]{zhu10a}. There is growing evidence that most FUors are embedded, as suggested by the observed continuum excess at $> 100 \ \mu\text{m}$ \citep{sandell01,green13} and by the recurrence timescale of FUor outbursts from the surveys \citep{scholz13,fischer19,contreraspena19}. Indeed, some FUors have been classified as Class 0/I objects \citep{kospal11,caratti11,Safron:2015a}.

Monitoring an embedded, episodically accreting source poses an observational challenge, as the protostellar emission is heavily attenuated by the surrounding envelope. The protostellar radiation is absorbed by the disk/inner envelope material very close to the protostar and re-emitted at longer wavelengths, leading to an increase in FIR and sub-mm emission with some delay due to light travel time \citep{johnstone13}.
Five embedded sources have sufficient modulation of long wavelength emission that indicate accretion bursts of at least a factor of a few: OO~Serpentis (a factor of 25 increase at $25 \ \mu\text{m}$; \citealp{kospal07}), NGC 6334l:MM1 (a factor of 4 increase at $1.3 \ \text{mm}$; \citealp{hunter17}), HOPS 383 (a factor of 35 increase at $70 \ \mu\text{m}$; \citealp{Safron:2015a}), S255IR NIRS3 (a factor of 2 increase at  $900 \ \mu\text{m}$; \citealp{liu18}), and EC 53 (a factor of 1.5 increase at $850 \ \mu\text{m}$; \citealp{yoo17}). 

To understand the observability of episodic accretion events in the embedded protostellar phase and the flux response at long wavelengths (70--1300 $\mu$m), \citet{macfarlane19a,macfarlane19b} performed radiative transfer calculations for early stages of young stellar objects (YSOs) with cavities, disks, and envelopes that were formed in 3D  hydrodynamic simulations.  For both FUor and EXor-sized accretion bursts, the internal increase in luminosity lead to more prominent brightening at sub-mm than at mm wavelengths.

Addressing the nature of episodic accretion in the embedded phase requires long term monitoring of star formation regions at (sub-) mm wavelengths. To this end, the JCMT Transient survey is currently undertaking monitoring of eight star forming regions within $500 \ \text{pc}$ of the Sun \citep{herczeg17}.  The primary aim of this survey is to observe continuum variability, which may relate to episodic accretion events in YSOs. The stability of the JCMT SCUBA-2 submillimeter camera provides a reliable measure of relative flux brightness changes to $2-3 \%$ for the brightest sources \citep{mairs17}. 

The JCMT Transient survey discovered that EC 53, a Class I YSO, exhibited an $850 \ \mu\text{m}$ flux increase by a factor of $1.5$, over a few months in 2016 \citep{yoo17}. Located in Serpens Main, EC 53 had previously showed a $2 \ \text{mag}$ brightening in the $K$-band, with periodicity of $\sim543 \ \text{days}$ \citep{hodapp12}.  Near-infrared (NIR) spectroscopy of EC 53 shows broad CO overtone absorption features, leading to a classification as a FUor candidate (Park et al.~in preparation). The JCMT observations closely follow the expected phase curve. \citet{yoo17} deduced that a protostellar luminosity enhancement by a factor $\sim4$ could recover the $850 \ \mu\text{m}$ flux increase, by using graybody components with different temperatures to account for the continuum emission at $850 \ \mu\text{m}$.  Variability
with smaller amplitudes has also been measured for several other protostars \citep{mairs17,johnstone18}.  

In this paper we model the emission from EC 53 both in quiescence and outburst phases in order to derive the enhancement factor in luminosity corresponding to the flux increase in the burst phase. We fit the emission from EC 53 using detailed radiative transfer modeling. 
The advantage of this approach is that we can determine how (a) the different components of the YSO and (b) the various sources of luminosity contribute to the observed flux. 

The paper is structured as follows. In \S~\ref{sec:rt_modeling} we outline the radiative transfer techniques adopted to model the dust continuum observations and density distributions used for the 2D and 3D modeling.
\S~\ref{sec:opacities} describes the adopted dust properties.  In \S~\ref{sec:param_seds} we find a fiducial 2D model that matches the Spectral Energy Distribution (SED) in the quiescent phase of EC 53 and explore the parameter space of YSO properties. \S~\ref{sec:3d_sed_fit} extends our analysis to explore the effect of the complex envelope density profile obtained in the 3D hydrodynamic simulation \citep{macfarlane19a,macfarlane19b}.
In \S~\ref{sec:burst} we present the response of the SED to an outburst using the fiducial models in the 2D and 3D. We also test the effect of external heating by the interstellar radiation field on our modeling results.
In \S~\ref{sec:rip} we analyze the 850 $\mu$m radial intensity profile and conclusions are presented in \S~\ref{sec:conclusions}.

\section{Radiative Transfer modeling of YSOs}\label{sec:rt_modeling}

We consider four components of an embedded YSO: the central protostar, circumstellar disk, envelope, and bipolar cavities. We also include external heating by the ambient radiation field.

\subsection{Radiative Transfer Method}\label{sec:rt_method}
We employ polychromatic radiative transfer (RT) using the software RADMC-3D \footnote{http://www.ita.uni-heidelberg.de/$\sim$dullemond/software/radmc-3d/} \citep{dullemond12} to perform simulations of different configurations that may represent EC 53. RADMC-3D uses the Monte Carlo Radiative Transfer (MCRT) method of \citet{bjorkwood01} to compute the equilibrium temperature for a density distribution and a set of luminosity sources. This is achieved by randomly emitting and subsequently propagating photon packets from each luminosity source through the computational domain. Once a photon is absorbed, its energy is deposited at that location, raising the local temperature. The photon at longer wavelengths is then re-emitted in a random direction. Once the equilibrium temperature has been calculated, the radiative transfer is solved and the wavelength dependent source function is calculated using a raytracing radiative transfer (RRT) algorithm. Therefore, we produce synthetic observations (SEDs and images). We assume a dust-to-gas ratio of 1:100 and photons pass through the model grids with isotropic scattering.

For the RT modeling in this paper, we adopt 436 pc as the distance to EC 53 in Serpens Main \citep{ortizleon17}. A foreground extinction of $A_{\rm V}$ = 9.6 mag is applied to the SEDs \citep{dunham15}.

\subsection{Radiation sources}\label{sec:rad_source}

The radiation emitted from the protostar and the background interstellar radiation field (ISRF) are considered. 

\subsubsection {Protostar}

Our fiducial protostellar model for EC 53 is assumed to be a blackbody with $6 \ \text{L}_\odot$, adopted from the extinction-corrected bolometric luminosity determined by \citet{dunham15}.
The protostellar luminosity during the outburst is a free parameter to fit the flux enhancement by a factor of $\sim1.5$ in the $850 \ \mu\text{m}$ flux. 

We adopt the stellar parameters of a typical T Tauri star, with a protostellar temperature of 4000 K and mass of 0.5 M$_{\odot}$. The effect of the stellar temperature on fluxes at sub-mm wavelengths is marginal because the sub-mm flux is proportional to the envelope temperature. The envelope temperature is determined mainly by the envelope optical depth, which in turn depends on the density distribution in the envelope and the protostellar luminosity. 
Although the outburst in EC 53 might be attributed to an unresolved binary \citep{hodapp12}, we assume only one internal luminosity source.

\subsubsection{ISRF}
\begin{figure}
   \centering
   {\includegraphics[trim={0.3cm 0.3cm 0.5cm 0.3cm},clip,width=85mm]{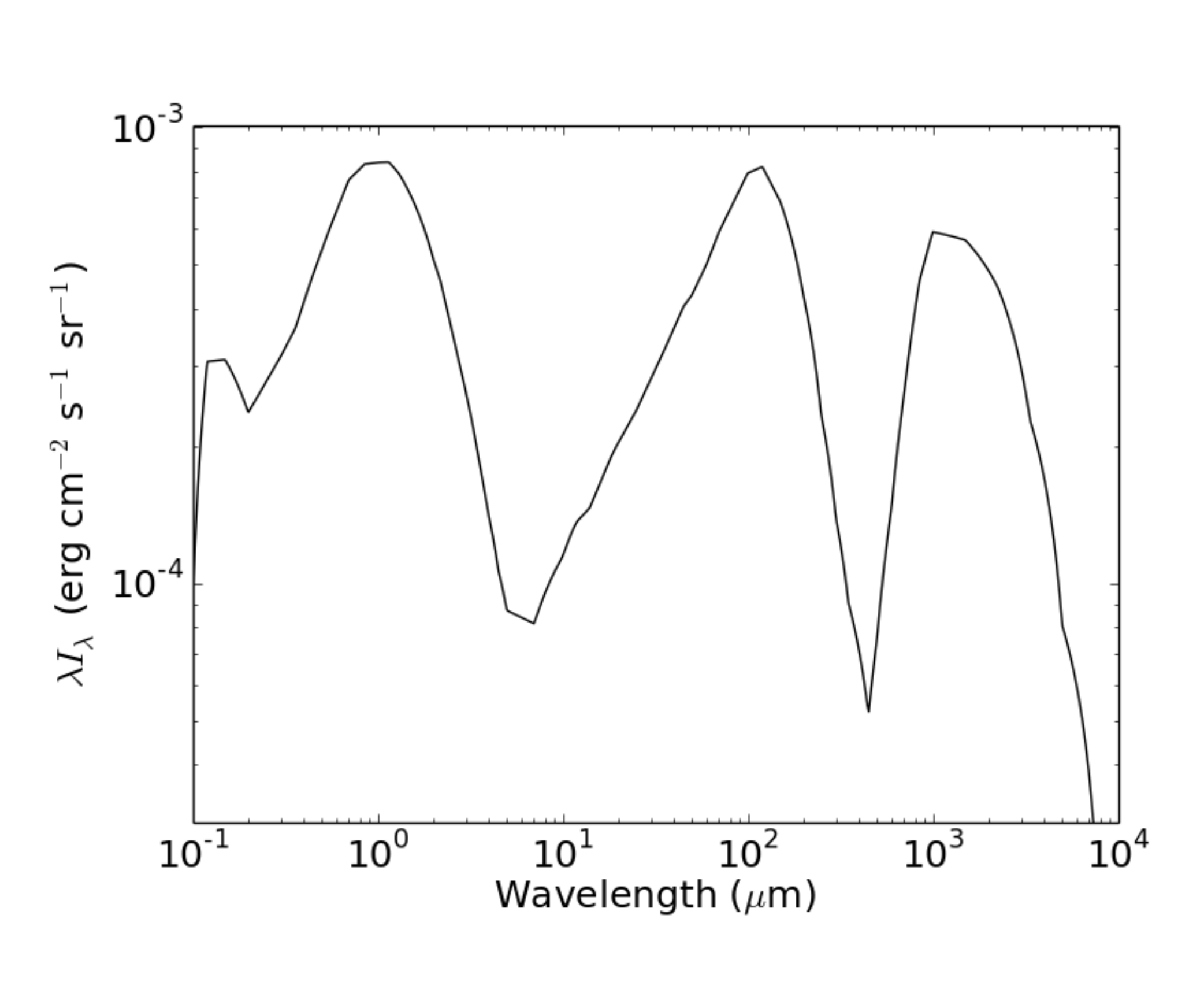}}
\caption{Intensity distribution of Black-Draine interstellar radiation field adopted in EC 53 modeling.}
\label{fig:isrf}
\end{figure}

We adopt the Black-Draine field \citep{evans01,je15}, which is the combination of two SEDs of ISRF: \citet{drain78} for $\lambda$ $<$ 0.36 $\mu$m and \citet{black94} for $\lambda$ $\geq$ 0.36 $\mu$m (see Figure~\ref{fig:isrf}).
In a deeply embedded protostellar system, the photoelectric heating induced by the ISRF is important for the temperature structure in the outer envelope \citep{evans01,lee04} because the photons from the central protostar cannot escape from the inner region due to the high optical depth.

\subsection{Density Field: 2D modeling}\label{sec:2d_model}

\begin{deluxetable*}{ccccc}
\tablecaption{2D Model Parameters\label{2dgrid}}
\tablehead{   \colhead{Parameter} & \colhead{Description} & \colhead{Values} & \colhead{Best Fit Model}  & \colhead{Parameter Use}}
\startdata
L (L$_{\sun}$) & Internal luminosity  & 6.0 & 6.0 &(fixed)\tablenotemark{a} \\
R$_{*}$(R$_{\sun}$) & Stellar radius    & 2.09 & 2.09  & (fixed)\\
T$_{*}$(K) & Stellar temperature        & 4000 & 4000  &(fixed)\\
M$_{*}$(M$_{\sun}$) & Stellar mass      & 0.5  & 0.5 &(fixed)\\
\hline
M$_{\rm disk}$(M$_{\sun}$) & Disk mass      & 0.0075 & 0.0075 & (fixed) \\
h(100AU) & Disk scale height at 100AU    & 48.0 & 48.0 & (fixed) \\
$\alpha$ & Disk radial density exponent      & 2.5   & 2.5  & (fixed) \\
$\beta$ & Disk flaring power  & 1.3 & 1.3 & (fixed) \\
R$_{disk,in}$(AU) & Disk inner radius                  & 0.34 & 0.34  & (fixed)\tablenotemark{a}\\
R$_{disk,out}$(AU) & Disk outer radius               & 90   & 90    & (fixed)\\
$\theta_{inc}$($^{\circ}$) & Disk inclination angle               & 30    & 30   & (fixed)\\
\hline
R$_{env,in}$(AU) & Envelope inner radius               & 0.34 & 0.34 & (fixed)\tablenotemark{a}\\
R$_{env,out}$(AU) & Envelope outer radius                    & 5,000, 10,000, 20,000 & 10,000 & \\
M$_{env}$(M$_{\sun}$) & Envelope mass           & 5.8 & 5.8 & (fixed)\\
$p$ & Envelope power-law index   & 1.0, 1.5, 2.0 & 1.5 & \\
\hline
b & Cavity shape exponent                      & 1.5   & 1.5  & (fixed) \\
$\theta_{cav}$($^{\circ}$) & Cavity opening angle & 10, 20, 30    & 20   &  \\
$\rho_{cav,in}$(g cm$^{-3}$) & Cavity inner density\tablenotemark{b}   & 1$\times$10$^{-19}$     & 1$\times$10$^{-19}$     & (fixed)\\
R$_{cav,bd}$(AU) & Cavity inner boundary radius\tablenotemark{c}   & 100     & 100     & (fixed)\\
\enddata
\tablenotetext{a}{For the quiescent phase.}
\tablenotetext{b}{The dust density of the cavity inside the R$_{cav,bd}$.}
\tablenotetext{c}{The radius at which the cavity density starts to decrease as $\rho \propto$ r$^{-2}$.}
\end{deluxetable*}

An axisymmetric density structure is adopted in a spherical coordinate system.
We use a logarithmic scale for $r$ and $\theta$  grid cells to deal with the protostellar system, which has a centrally concentrated density structure and a size covering $\sim$ 5 orders of magnitude in radius (from 0.1 AU to 10,000 AU). 
Logarithmic spacing gives better spatial resolution near the protostar and disk midplane, where the density is very high. This method is effective to keep the photons from being trapped in the optically thick cells.
The density distributions of YSO components and their parameterization are as below: 

\paragraph*{Protostellar disk. \ } In YSOs, material infalls from the envelope to the disk, which is a temporary mass reservoir before the material accretes to the protostar. 
We use a standard flared accretion disk for the disk density structure \citep{shakura73,lynden-bell74,hartmann98}, given by 
\begin{eqnarray}
\label{eq:disk_den}
\rho \left ( \varpi ,z \right ) = \rho_{disk,0}\left ( 1-\sqrt{\frac{R_{*}}{\varpi}}\right )\left (\frac{R_{*}}{\varpi } \right )^{\alpha}exp\left \{ -\frac{1}{2}\left [ \frac{z}{h} \right ]^{2} \right \},
\end{eqnarray}
where $\rho_{disk,0}$ is the normalization constant calculated from the integral of the density over the entire disk, which is equal to the disk mass. $\varpi$ is the cylindrical radius ($\varpi =  \sqrt{x^{2}+y^{2}}$). The disk scale height $h$ is defined as $h\propto \varpi ^{\beta }$, where $\beta$ is the disk flaring power. $R_{*}$ is the stellar radius and $\alpha$ is the radial density exponent. 
In this model we adopt 2.5 and 1.3 for $\alpha$ and $\beta$, respectively, to describe the flared disk in an early stage of star formation \citep{whitney03,tobin13}.
The inner radius of the dust disk is 0.34 and 0.58 AU for quiescent and outburst phases, respectively, inside of which the dust temperature is above the dust destruction temperature of $\sim$1200 K \citep{macfarlane19a}.

We divide the disk into two zones depending on the density. One is the dense disk midplane (n$_{\rm{H2}}$ $>$ 10$^{10}$ cm$^{-3}$) and the other is disk atmosphere. We apply different dust properties for the two zones (see Section \ref{sec:opacities}).

\paragraph*{Envelope. \ }A simple power-law density profile is adopted for the envelope;
\begin{eqnarray}
\label{eq:envelope}
\rho(r) = \rho_{env,0}\left(\frac{r}{R_{env,in}} \right)^{-p}, 
\end{eqnarray}
where $R_{env,in}$ is the envelope inner radius and $p$ is the power-law index. We set $R_{env,in}$ as the same as the disk inner radius since envelope material could infall onto the disk \citep{terebey84}.  We calculate models for density profiles with power-law indices $p$ of 1.0, 1.5 and 2.0. Generated density profiles are extended to the 2D grid space.
$\rho_{env,0}$ is the density at the $R_{env,in}$ and scaled to have a total envelope mass M$_{env}$ of $\sim$ 5.8 M$_{\odot}$ (see  \S~\ref{sec:ob_constraints}). $R_{env,out}$ is the envelope outer radius.
Since the $R_{env,in}$, $R_{env,out}$ and M$_{env}$ are fixed, $\rho_{env,0}$ varies considerably among models with different density power-law indices.

\paragraph*{Bipolar outflow. \ } As a consequence of the mass accretion to the central protostar, ionized jets, outflows and high velocity winds emerge and sweep out the envelope material, leaving bipolar cavities orthogonal to the disk.
Through the cavities, the radiation from the central protostar escapes directly or by scattering, which contributes significantly to the fluxes in the NIR and mid-infrared (MIR) in the emerging SED. 
We include a curved cavity structure from the central star to the envelope using the equation,
\begin{eqnarray}
\label{eq:cavity}
\textit{z} = c\varpi ^{d}, 
\end{eqnarray}
where $c$ is  $R_{env,out}/\left (R_{env,out} \rm{tan}\theta _{\rm cavity}\right )^{1.5}$ with the envelope outer radius of R$_{env,out}$, the cavity opening angle of $\theta _{\rm cavity}$, and the cavity shape exponent of $d$. We adopt d=1.5, following previous fits to the SEDs of other Class 0/I objects \citep{tobin13,je15}. The cavity opening angle is defined as the angle between the edge of the cavity and the zenith of the system.
Following the convention from \citet{yang17}, we set a constant dust density 10$^{-19}$ g cm$^{-3}$ within 100 AU and the density decreases as $\rho\propto r^{-2}$ beyond 100 AU.
The parameters used in the 2D model are summarized in Table \ref{2dgrid}.

\subsection{Density Field: 3D modeling}\label{sec:3d_model}
\subsubsection{Hydrodynamic simulation}

To further investigate the importance of possible asymmetries in the envelope, we adopt the output of a smoothed particle hydrodynamic (SPH) simulation for the formation of a YSO \citep{macfarlane19a,macfarlane19b} from the collapse of a pre-stellar core \citep{stamatellos12} with the mass of $5.4 \ \text{M}_\odot$ and the outer radius of $R = 50,000 \ \text{AU}$ for which the initial density profile of the cloud is Plummer-like \citep[see][]{stamatellos12}. The resulting density profile is rescaled for consistency between the 2D and 3D models to have the radius of 10,000 AU and total envelope gas mass of 5.8 M${_\odot}$  (Section \ref{sec:ob_constraints}). 

\begin{figure*}
   \centering
   \subfigure{\includegraphics[trim={0.1cm 0.1cm 0.1cm 0.1cm},clip,width=0.98\columnwidth,keepaspectratio]{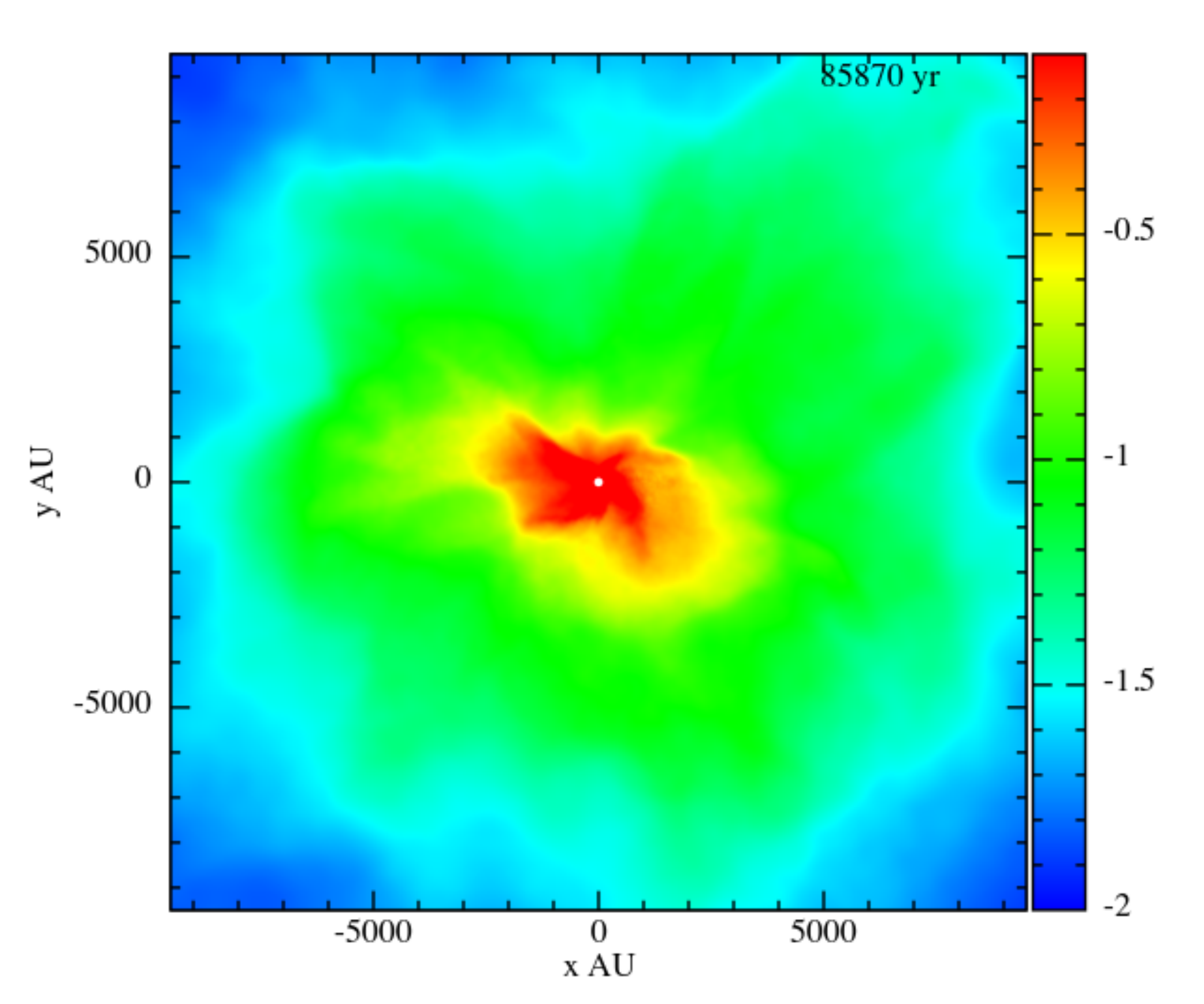}}
   \subfigure{\includegraphics[trim={0.1cm 0.1cm 0.1cm 0.1cm},clip,width=0.98\columnwidth,keepaspectratio]{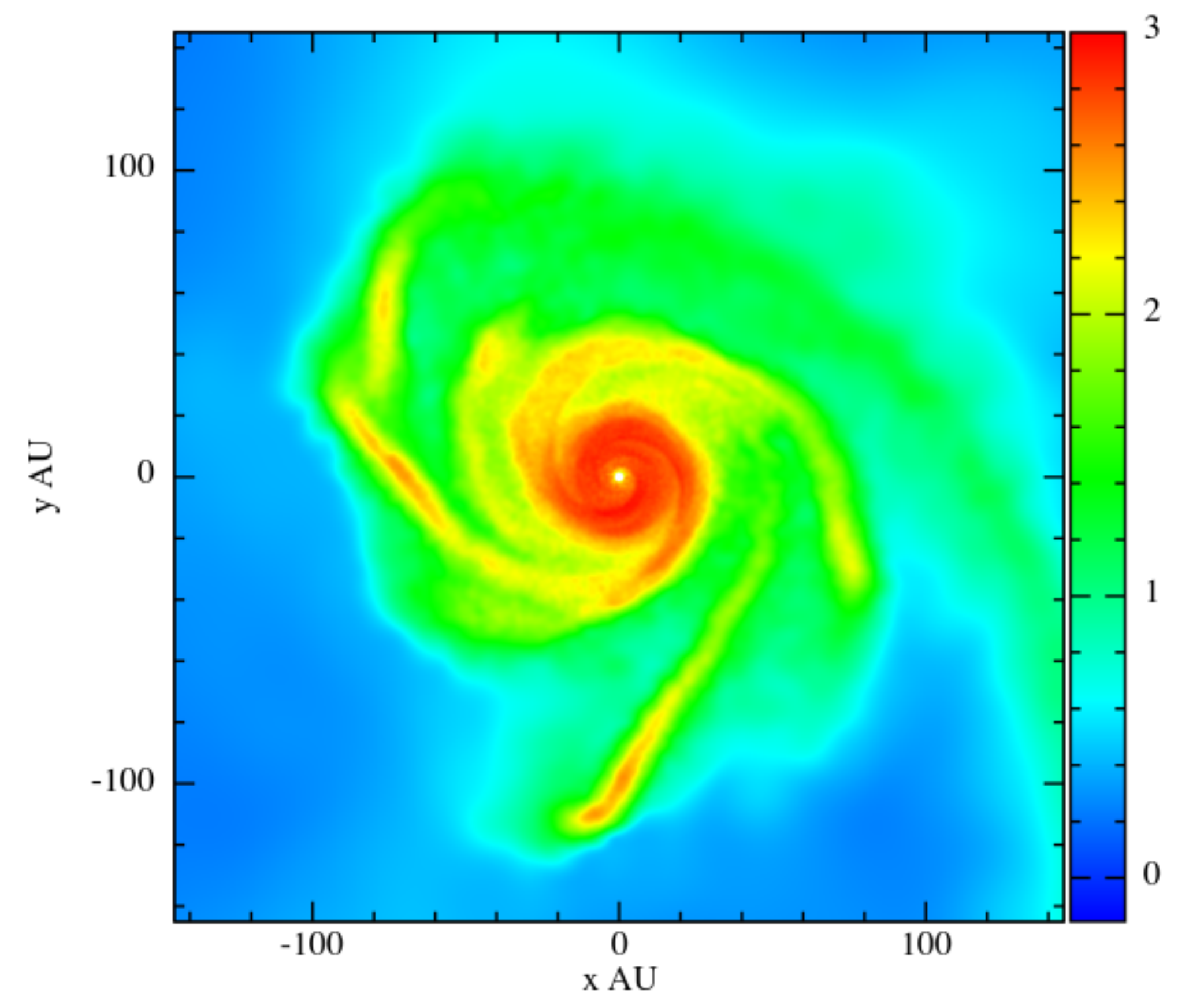}}
\caption{Density field: 3D modeling; Column density map (logarithmically scaled ; $\text{g cm}^{-2}$) of the  snapshot used for modeling EC 53. The left panel shows the large scale structure, whereas the right panel shows the central region of the YSO.}
\label{fig:3d_col_dens}
\end{figure*}

We choose a snapshot from this simulation to represent the density structure of EC 53.  The column density maps of the snapshot (Figure \ref{fig:3d_col_dens}) show both the large scale (left) and the small scale (right) structure of the YSO. For simplicity, we use the same snapshot both for the quiescent and outbursting phase of EC 53. Due to the outburst, the structure of the very innermost regions of the YSO may be altered, however, the structure of the outer envelope that produces sub-mm emission is not expected to change \citep{macfarlane17}. The 2D presented in Section \ref{sec:burst}  follows the same approach.
The density profile of the envelope in this simulation snapshot is approximately $\rho(r) \propto r^{-2}$. Toward the inner regions of the YSO (i.e. near the protostellar disk), the radial density profile is even steeper, i.e.  corresponds to a very centrally condensed YSO.

\subsubsection{Construction of the Radiative Transfer Grid}

Since RADMC-3D uses a grid-based approach for all numerical computations, we translate the 3D SPH density distribution to a grid, following the approach of \citet{macfarlane19a,macfarlane19b}. 
The computational domain is subdivided into a set of 8 equal volume cubes (octants) and we continue this until each sub-octant hosts $\leq5$ SPH particles, or until a maximum refinement level of $20$ is reached. We then calculate the density at each octant using the mass of the particles that contribute to the mass of the octant.
Finally, from the gas density distribution a dust density is calculated using a dust-to-gas ratio of 1:100.

To account for the destruction of dust by either sublimation or sputtering \citep{lenzuni95,duschl96} we include a dust cavity near the protostar. The dust destruction radius is determined by the destruction temperature of 1200 K. 
The density inside the dust destruction radius is  set to zero \citep{macfarlane19a,macfarlane19b}.

\subsubsection{Definition of the YSO components}

\paragraph*{Bipolar outflow. \ }

The SPH simulation does not account for bipolar outflows, so we seed these in the density field. We adopt the prescription of Equation~\ref{eq:cavity}, with parameters and density profile equivalent to the 2D modeling as described in \S~\ref{sec:2d_model}.

\paragraph*{Protostellar disk. \ }

The protostellar disk forms self-consistently in the simulation. To calculate the radial extent, first we calculate the disk midplane from the angular momentum vector of the accreting protostar, assuming that most of the accreted mass comes from the disk. Then we define the radial disk extent as the radius at which the azimuthally averaged surface density falls below $20 \ \text{gcm}^{-2}$. This definition gives a disk extent to the Keplerian disk radius \citep{stamatellos12, macfarlane17}. For the snapshot taken for EC 53, the disk radius is $ r_{xy} = 115 \ \text{AU}$.  

We use different dust compositions for the disk midplane and the disk atmosphere (see Section \ref{sec:opacities}), with a boundary between these two layers of one scale height, $h(r_{xy})= c_{s}/\Omega$, where  $c_{s}$ is the sound speed and $\Omega$ is the Keplerian angular velocity. The vertical extent of the disk atmosphere region is set to $3 h$. 

\paragraph*{Envelope. \ } The envelope region is defined as any location not within the cavity nor within either disk component. 
\subsection{Opacities of the different YSO components}\label{sec:opacities}
\begin{figure}
   \centering
     {\includegraphics[trim={0.3cm 0.3cm 0.5cm 0.3cm},clip,width=85mm]{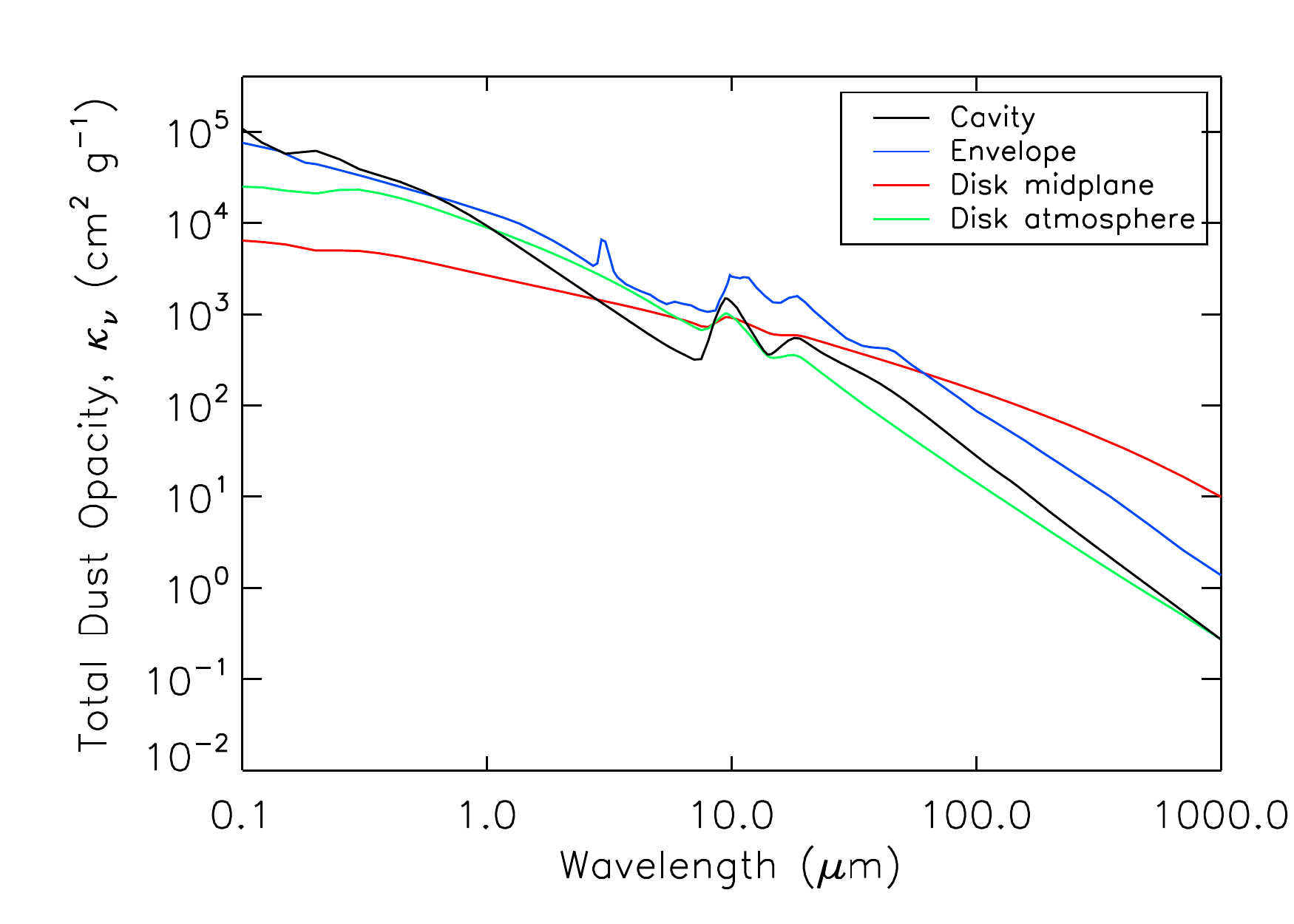}}
\caption{Total dust opacity, $\kappa_\nu$, of opacity tables adopted in modeling of EC 53. Black, blue, red and green lines represent opacities of the bipolar outflow cavity \citep{kim94}, the envelope \citep{oh94}, the disk midplane \citep{wood02}, and the disk atmosphere \citep{cotera01}, respectively.}
\label{fig:opacities}
\end{figure}

We adopt different opacities for the bipolar outflow cavity, envelope, disk midplane, and disk atmosphere. 
For the envelope, we adopt the opacity from the 5th column of Table 1 in \citet{oh94} (OH5), which represents the grains growing by the coagulation and accretion of thin ice mantles for 10${^5}$ yrs at a density of 10${^6}$ cm$^{-3}$. 
For the outflow cavity, we take the grain size distribution from \citet{kim94}, which is similar to the ISM dust in Taurus, but smaller than dust grains in the YSO envelopes. 
For the disk, we apply two different dust properties.
At the dense disk midplane, we use a large grain model presented by \citet{wood02}. For the less dense region, a grain model presented by \citet{cotera01} is used; the grains of this model are larger than the ISM grain and smaller than those in disk midplane.
The total opacities (absorption plus scattering) as a function of both the wavelength and regions are plotted in Figure~\ref{fig:opacities}. 

\section{The quiescent-phase SED of EC 53}\label{sec:param_seds}
\subsection{2D modeling: Quiescent phase SED}\label{sec:2d_sed_fit}

The model that fits the SED of EC 53 in the quiescent phase is considered as our fiducial model, from which we explore various physical parameters.

\subsubsection{Observational constraints}\label{sec:ob_constraints}
The photometric data for EC 53 were collected from the literature and archive; we adopt photometry from 2MASS J, H, K$_{\rm s}$ bands \citep{cutri03}, Spitzer Infrared Array Camera 3.6--8.0 $\mu$m (IRAC; \citealt{fazio04}) and Multiband Imaging Photometer 24--70 $\mu$m (MIPS; \citealt{rieke04}) data by {\it Spitzer Space Telescope} \textquotedblleft cores to disks" (c2d) and Gould Belt surveys \citep{dunham15}, Wide-field Infrared Survey Explorer (WISE) 12 and 22 $\mu$m \citep{wright10,mainzer14}, and CSO/SHARC-II 350 $\mu$m and Bolocam 1.1 mm data from \citet{dunham15}. 
We also include {\it Herschel}/PACS 70, 100 and 160 $\mu$m data \citep{marton17}, which is obtained from the IRSA archive\footnote{http://irsa.ipac.caltech.edu/}.

For the 850 $\mu$m flux of EC 53, we adopt the measurement for faint and bright phases from \citet{yoo17}. 
For the fiducial model, we use the envelope mass of 5.8 M$_{\sun}$, which is calculated with the
1.1 mm flux given by \citet{dunham15} and the distance of 436 pc \citep{ortizleon17}.
The adopted photometric data is summarized in Table \ref{tab:obs}.

\begin{deluxetable}{ccccc}
\tablecaption{Photometry table \label{tab:obs}}
\tabletypesize{\scriptsize}
\tablehead{
\colhead{Wavelength[$\mu$m]} & \colhead{Flux$_{\rm total}$[mJy]}  & \colhead{Instrument} & \colhead{Reference}}
\startdata
1.25	& 0.33$\pm$0.03&   2MASS/J & 1 \\
1.25 & 0.39$\pm$0.01&   UKIRT/J & q  \\
1.25 & 1.56$\pm$0.03&   UKIRT/J & o  \\
1.65	& 0.92$\pm$0.12 & 2MASS/H & 1 \\
1.65 & 1.99$\pm$0.03&   UKIRT/H & q  \\
1.65 & 8.80$\pm$0.14&   UKIRT/H & o  \\
2.17	& 4.30$\pm$0.29 &   2MASS/K$_{\rm s}$ & 1 \\
2.2 & 8.03$\pm$0.12&   UKIRT/K & q  \\
2.2 & 35.40$\pm$0.49&   UKIRT/K & o \\
3.4 & 13.55$\pm$0.25&   WISE/W1 & q \\
3.4 & 58.82$\pm$5.08&   WISE/W1 & o \\
3.6	& 31.0$\pm$3.00 &   $Spitzer$/IRAC & 1 \\
4.5	& 73.0$\pm$4.90&   $Spitzer$/IRAC &1 \\
4.6 & 50.83$\pm$0.66&   WISE/W2 & q \\
4.6 & 244.68$\pm$7.78&   WISE/W2 & o \\
5.8	& 140.0$\pm$7.20 &   $Spitzer$/IRAC &1\\
8.0	& 210.0$\pm$11.0 &  $Spitzer$/IRAC & 1 \\
12	& 180.0$\pm$2.50&   WISE/W3 &1 \\
12	& 180.0$\pm$3.50 &   WISE/W3 & q \\
22	& 950.0$\pm$17.0 &   WISE/W4 &1 \\
22	& 1028.8$\pm$7.80 &   WISE/W4 & q \\
24	& 990.0$\pm$100.0 &  $Spitzer$/MIPS & 1\\
70	& 8500.0$\pm$910.0 &  $Spitzer$/MIPS & 1\\
70	& 10817.3$\pm$603.1&   $Herschel$/PACS & 2 \\
100	& 13597.3$\pm$632.8&   $Herschel$/PACS & 2 \\
160	& 20985.9$\pm$8110.5 &   $Herschel$/PACS & 2\\
350	& 20700.0$\pm$5400.0 &   CSO/SHARC-II& 1  \\
850	& 2875.0$\pm$280.0 &   JCMT/SCUBA-2 & 3  \\
1100	& 1500.0$\pm$150.0 &   CSO/Bolocam & 1 \\
\enddata
\tablenotetext{1}{\citealt{dunham15}}
\tablenotetext{2}{IRSA archive: http://irsa.ipac.caltech.edu/} 
\tablenotetext{3}{\citealt{yoo17}}
\tablenotetext{q}{Quiescent phase}
\tablenotetext{o}{Outburst phase}
\end{deluxetable}

\subsubsection{A fiducial model and parameter exploration}\label{sec:fid}

\begin{figure}
   \centering
   \subfigure{\includegraphics[trim={0.5cm 0.5cm 0.5cm 0.5cm}, clip, width=0.98\columnwidth, keepaspectratio]{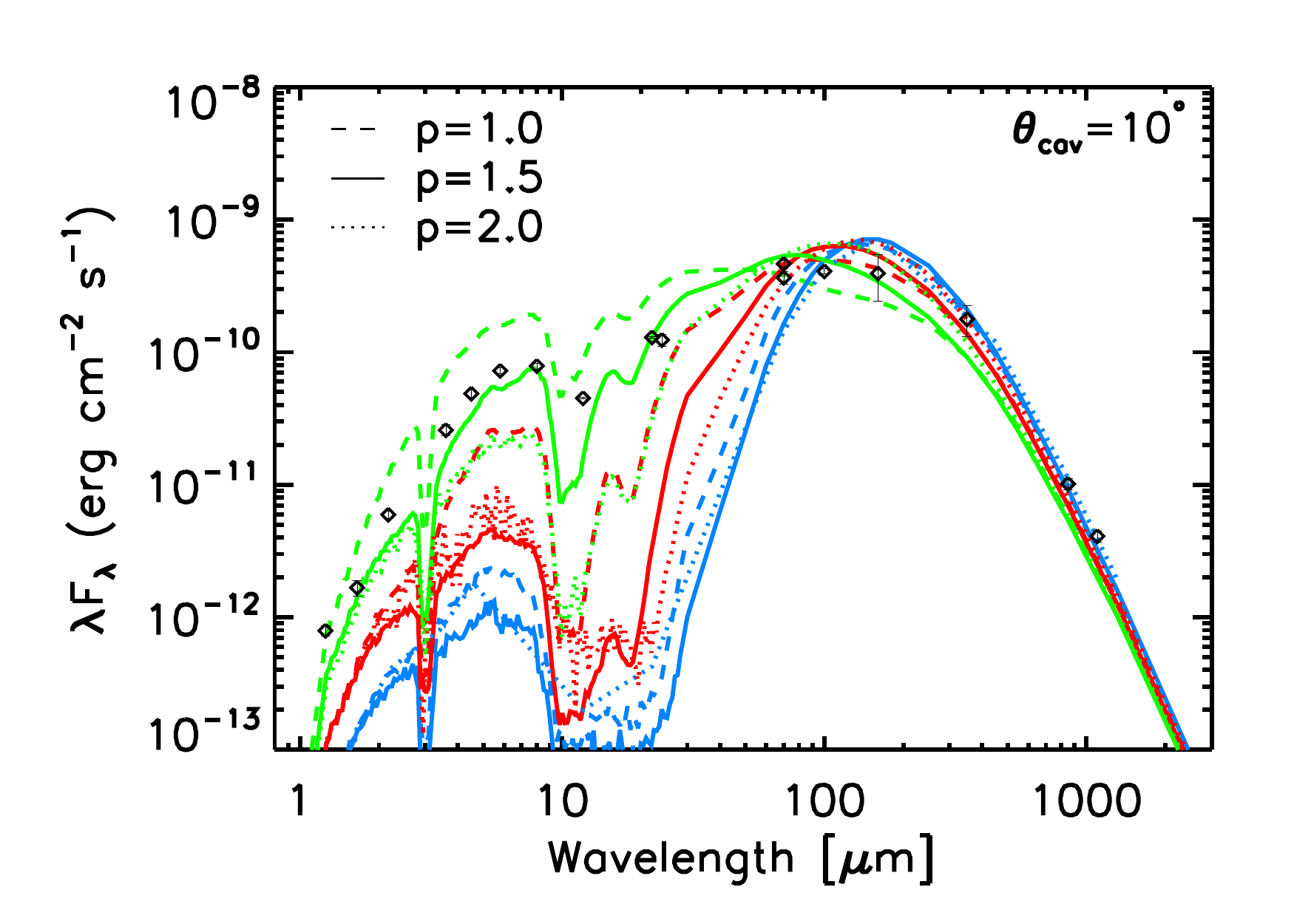}}
   \subfigure{\includegraphics[trim={0.5cm 0.5cm 0.5cm 0.5cm}, clip, width=0.98\columnwidth, keepaspectratio]{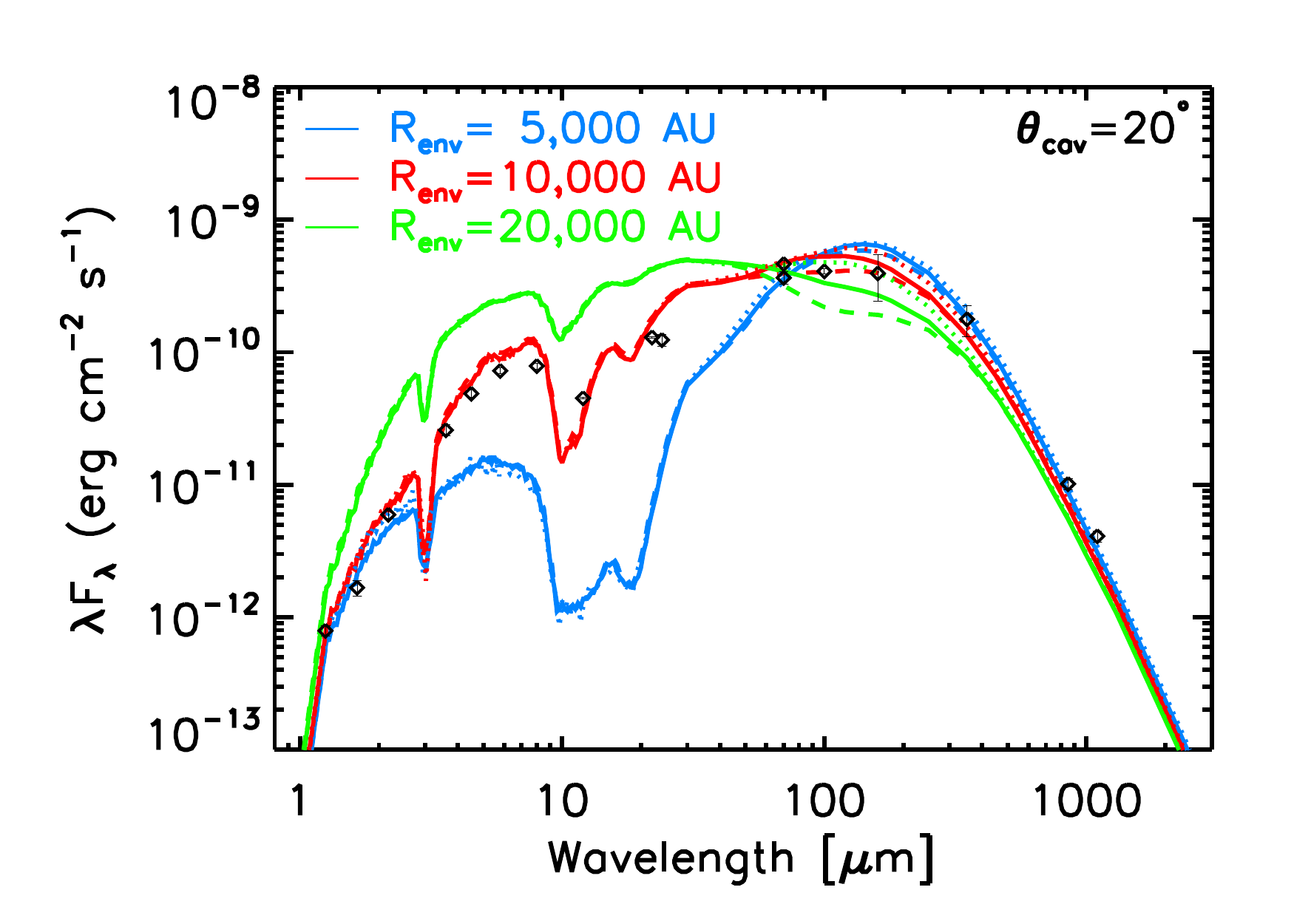}}
   \subfigure{\includegraphics[trim={0.5cm 0.5cm 0.5cm 0.5cm}, clip, width=0.98\columnwidth, keepaspectratio]{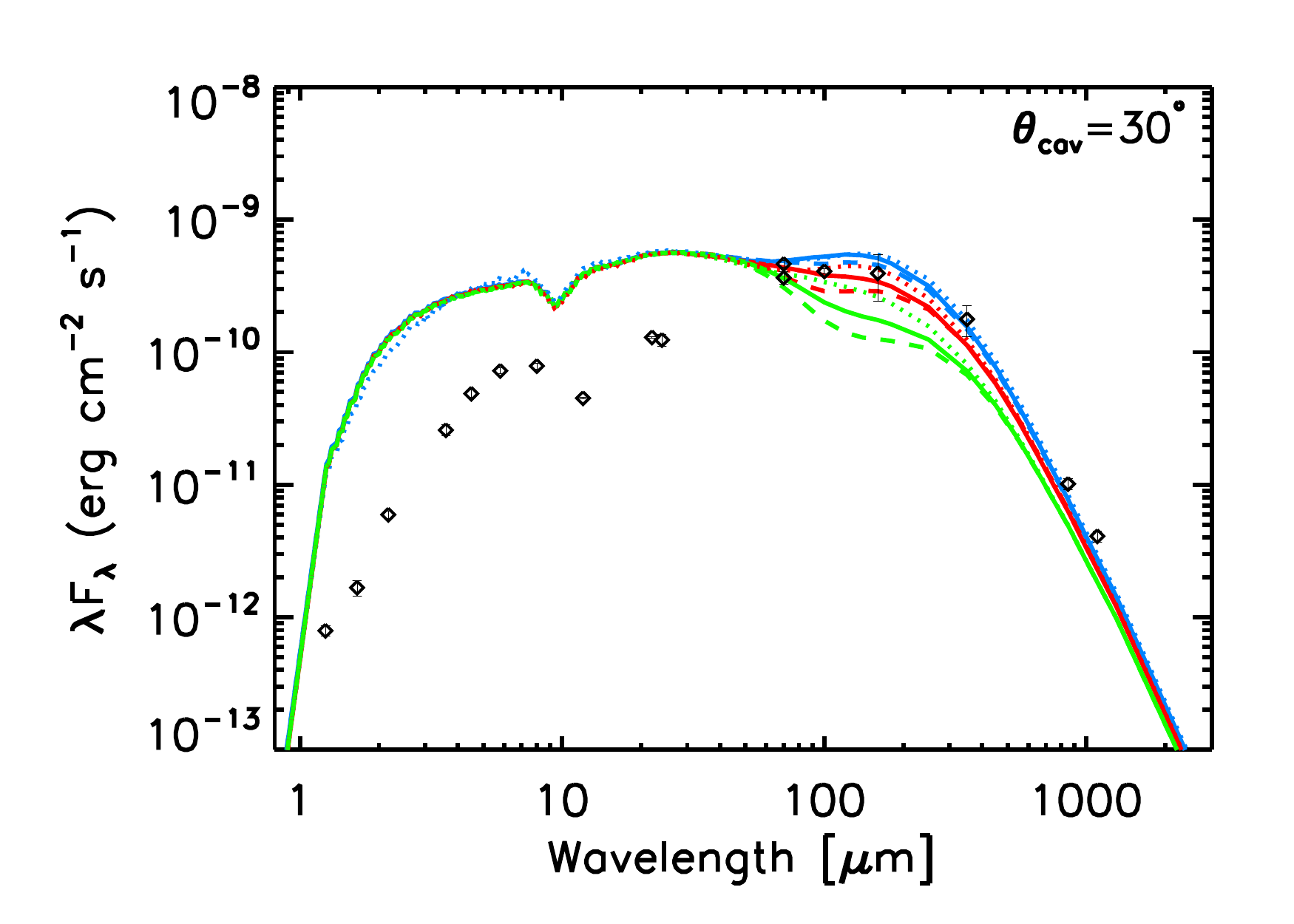}}
\caption{27 cases of the 2D modeling. Top, middle, and bottom panels show the models with cavity opening angles of $\theta$ = 10, 20 and 30$^{\circ}$ in Equation \ref{eq:cavity}, respectively. Each panel presents the models with varying three different power-law indices ($p$ = 1.0, 1.5 and 2.0 in Equation \ref{eq:envelope} in different line types, and envelope size (R = 5,000, 10,000, and 20,000 AU) in different colors. The black diamonds represent the observed fluxes.}
\label{fig:2d_27cases}
\end{figure}

\begin{figure}
   \centering
   \subfigure{\includegraphics[trim={0.1cm 0.3cm 3.0cm 0.1cm}, clip, width=0.49\columnwidth, keepaspectratio]{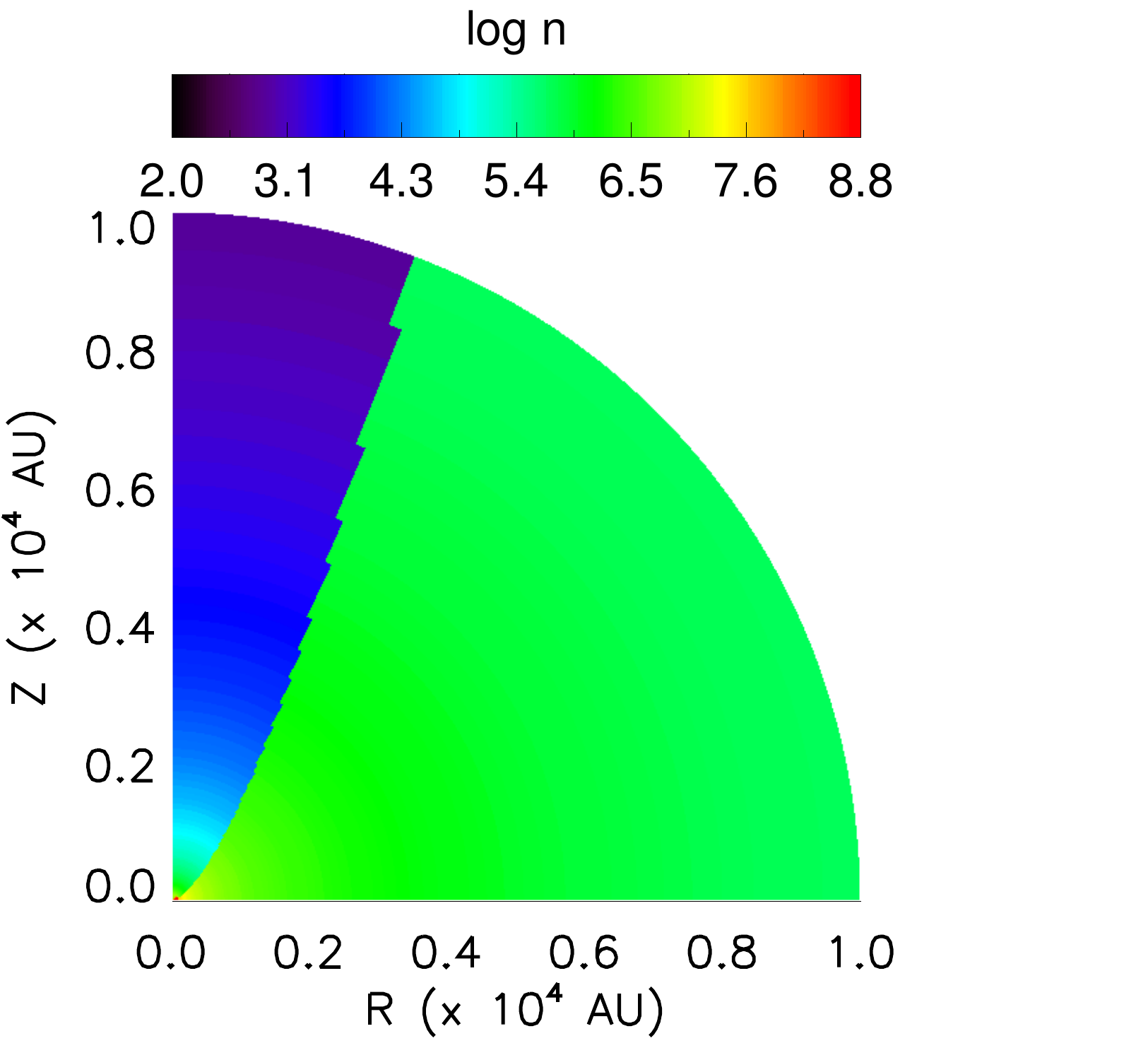}}
   \subfigure{\includegraphics[trim={0.1cm 0.3cm 3.0cm 0.1cm}, clip, width=0.49\columnwidth, keepaspectratio]{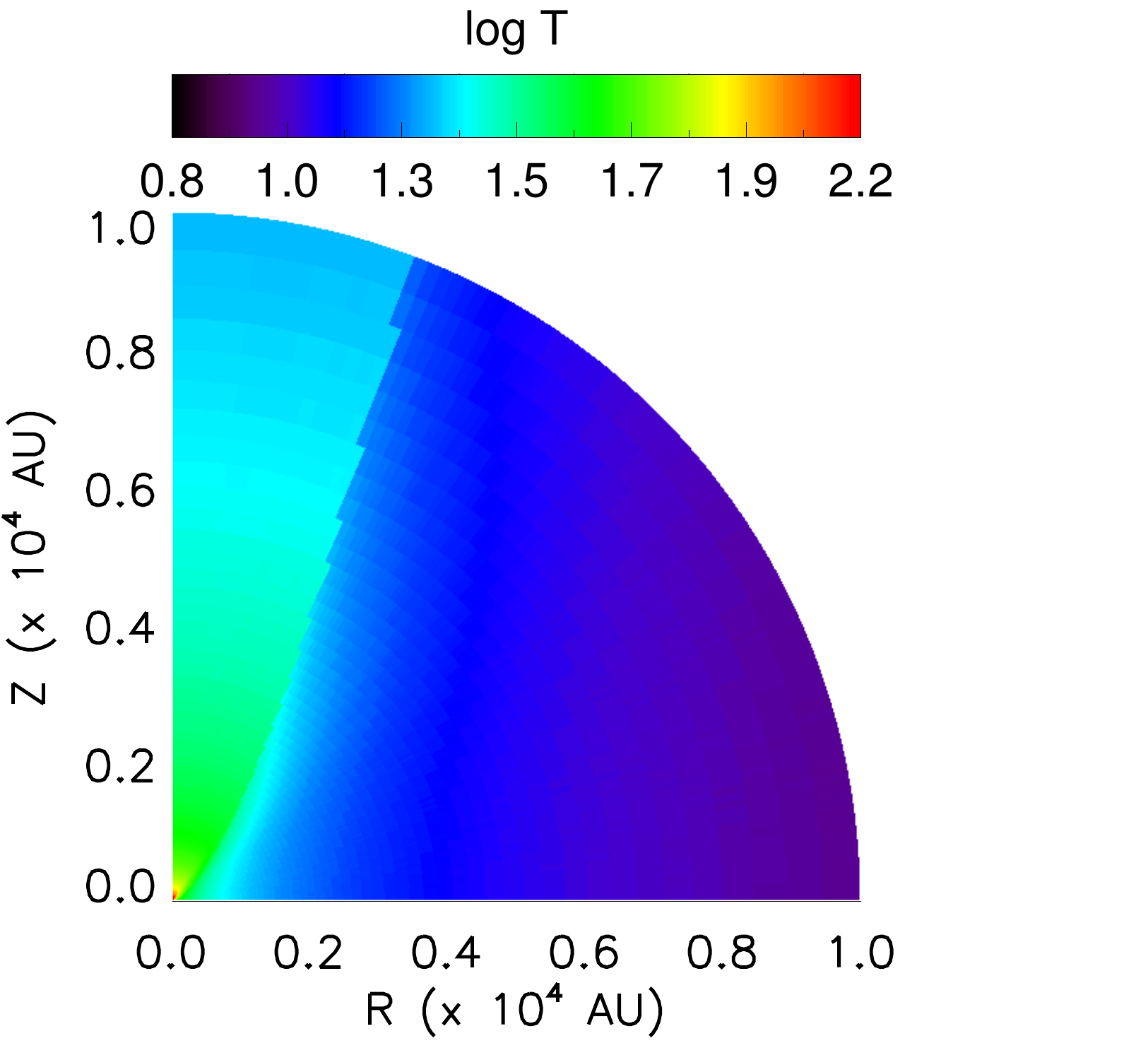}}\\
    \subfigure{\includegraphics[trim={0.1cm 0.3cm 3.0cm 0.1cm}, clip, width=0.49\columnwidth, keepaspectratio]{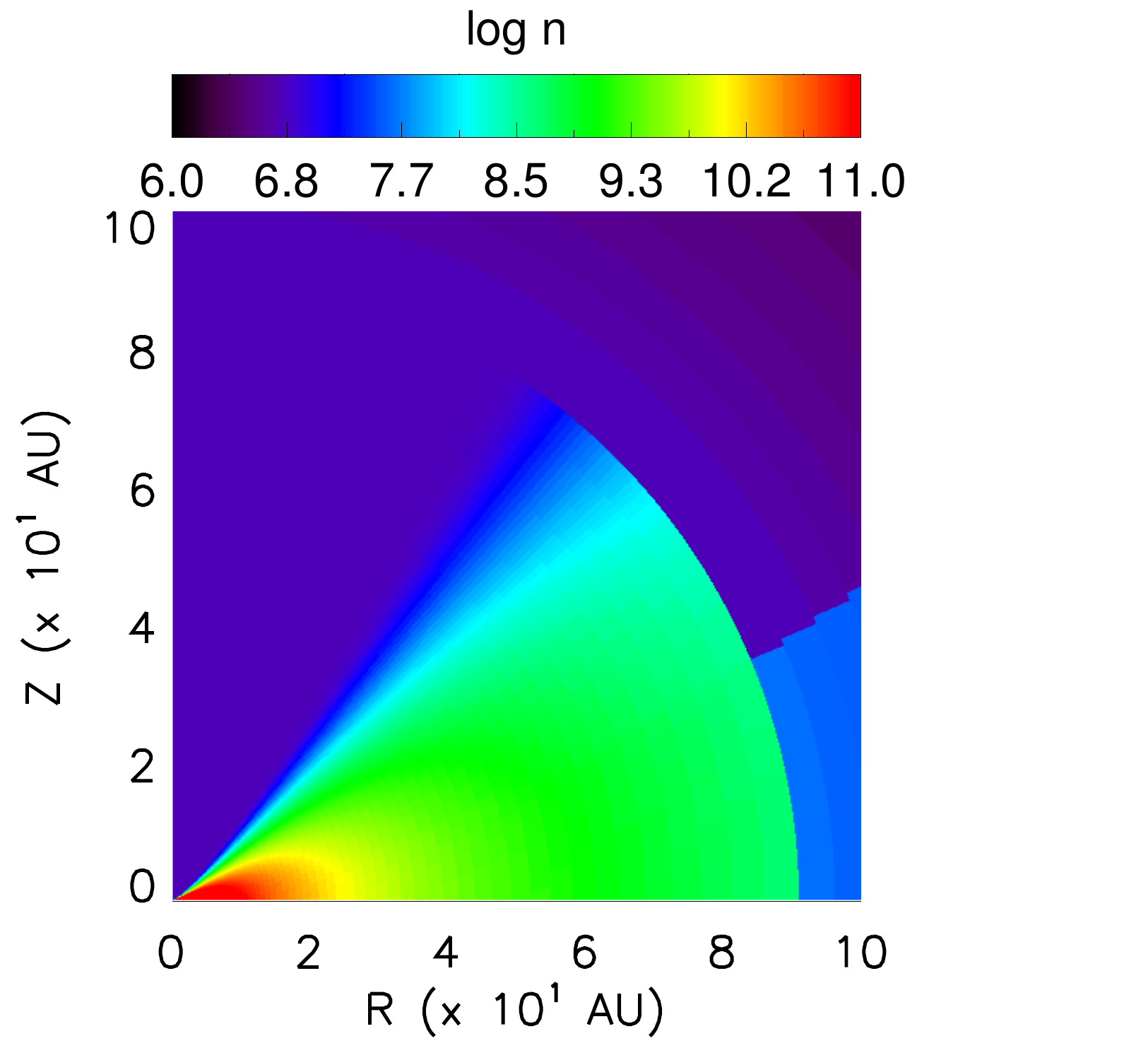}}
   \subfigure{\includegraphics[trim={0.1cm 0.3cm 3.0cm 0.1cm}, clip, width=0.49\columnwidth, keepaspectratio]{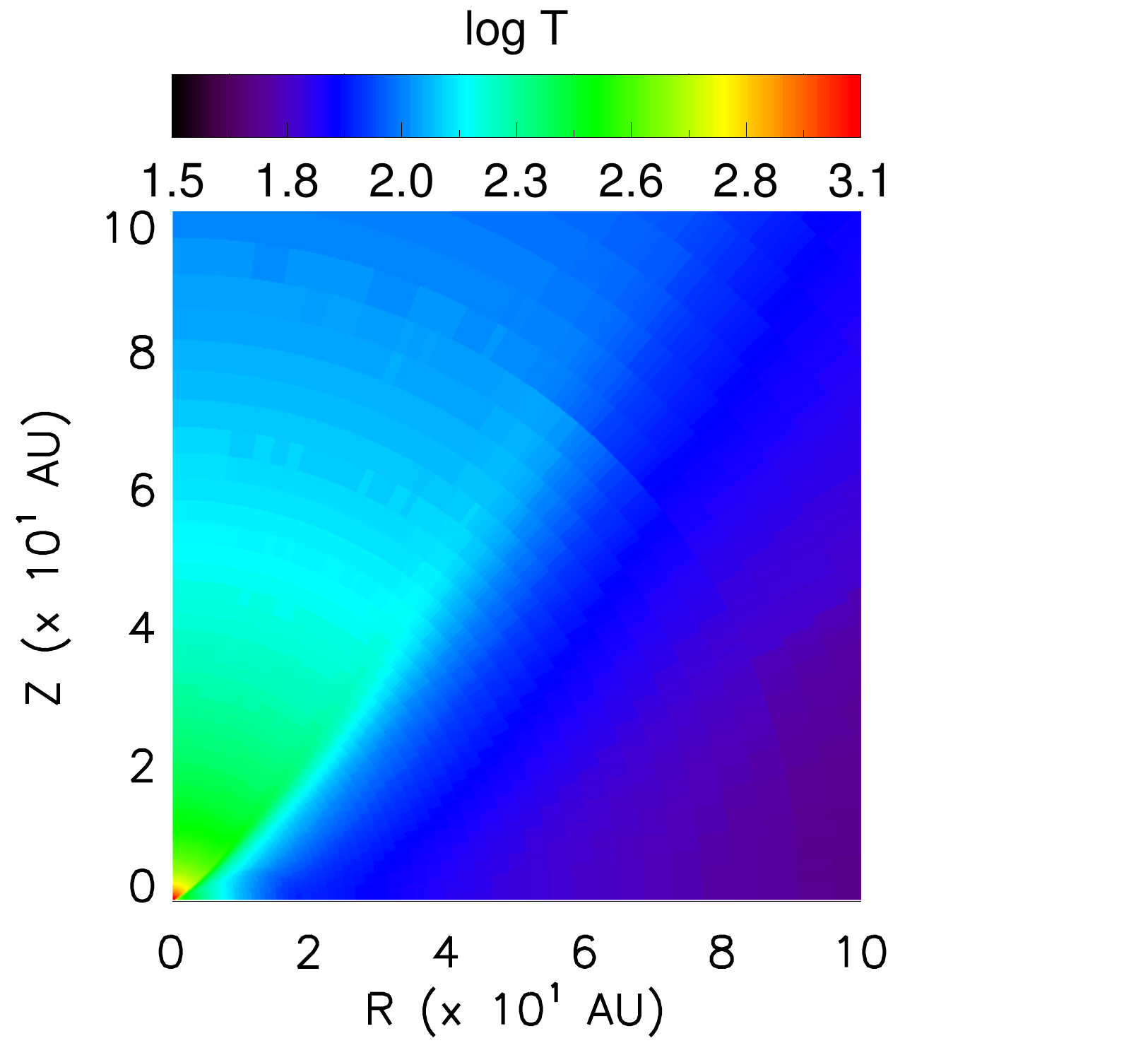}}
\caption{The density (left) and temperature (right) distributions of the 2D fiducial model. Top panels show overall distributions and bottom panels show the distributions up to $\sim$100 AU. In each panel, central protostar is located at (0.0, 0.0) position.}
\label{fig:2d_T_den}
\end{figure}

We explore the effects of envelope density structure, cavity opening angle and envelope size on SED when the protostar is the sole luminosity source.
Figure \ref{fig:2d_27cases} shows 27 SED models with different physical parameters at the fixed disk inclination of 30$^{\circ}$, which fits the observed SED the best. The disk inclination angle $i$ is defined as an angle between the plane of the sky and the plane of the disk: $i$ = 0$^{\circ}$ for a face-on disk and $i$ = 90$^{\circ}$ for a edge-on disk. 
We test envelope density distributions with three different power-law indices ($p$ = 1.0, 1.5 and 2.0 in Equation \ref{eq:envelope}),  cavity opening angles ($\theta$ = 10, 20 and 30$^{\circ}$ in Equation \ref{eq:cavity}), and envelope sizes (R = 5,000, 10,000, and 20,000 AU). 
We consider both the SED and the radial intensity profile of the 850 $\mu$m image (see Section \ref{sec:rip}) to find the best-fit model based on the $\chi^{2}$ minimization method. 
The best-fit model of EC 53 in the quiescent phase is the model with an envelope density power-law index of 1.5, a cavity opening angle of 20$^{\circ}$, and an envelope size of 10,000 AU, as presented in the solid red line in the middle panel of Figure \ref{fig:2d_27cases}. {\bf The $\chi^{2}_{red}$ value (Equation 6 in \citealt{robitaille07}) for the best-fit SED model, which is further constrained by the radial intensity profile at 850 $\mu$m, is 18.25. This high $\chi^{2}_{red}$ value is mainly due to the short wavelength regime, where the disk emission contributes the most. For a better fit with a lower $\chi^{2}_{red}$ value, the disk properties need to be constrained more precisely, which is beyond the scope of this work.} We set this model as our fiducial model. The density and temperature distributions of the fiducial model are shown in Figure \ref{fig:2d_T_den}.

\begin{figure}
   \centering
   \subfigure{\includegraphics[trim={0.5cm 0.5cm 0.5cm 0.5cm}, clip, width=0.98\columnwidth, keepaspectratio]{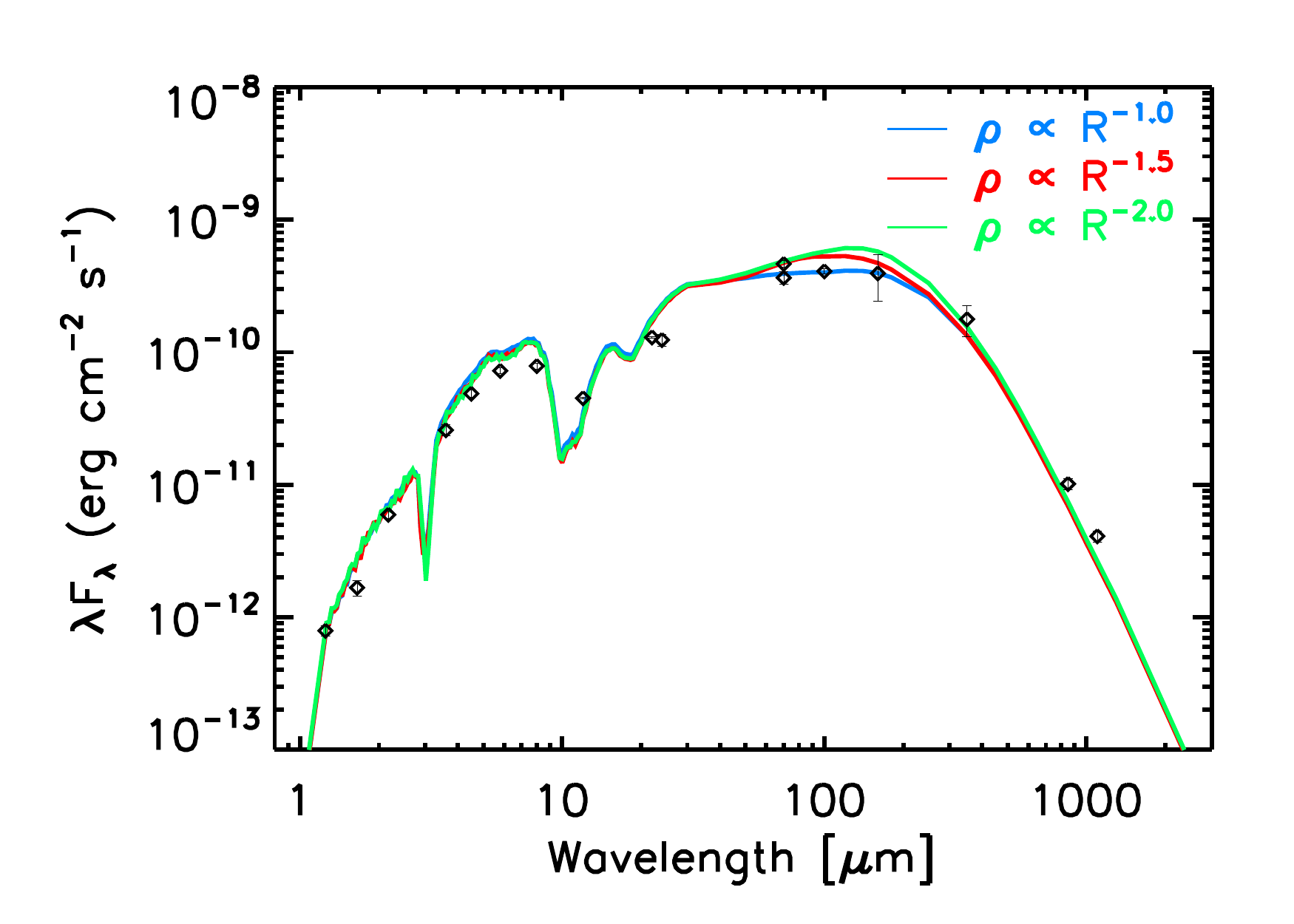}}
   \subfigure{\includegraphics[trim={0.5cm 0.5cm 0.5cm 0.5cm}, clip, width=0.98\columnwidth, keepaspectratio]{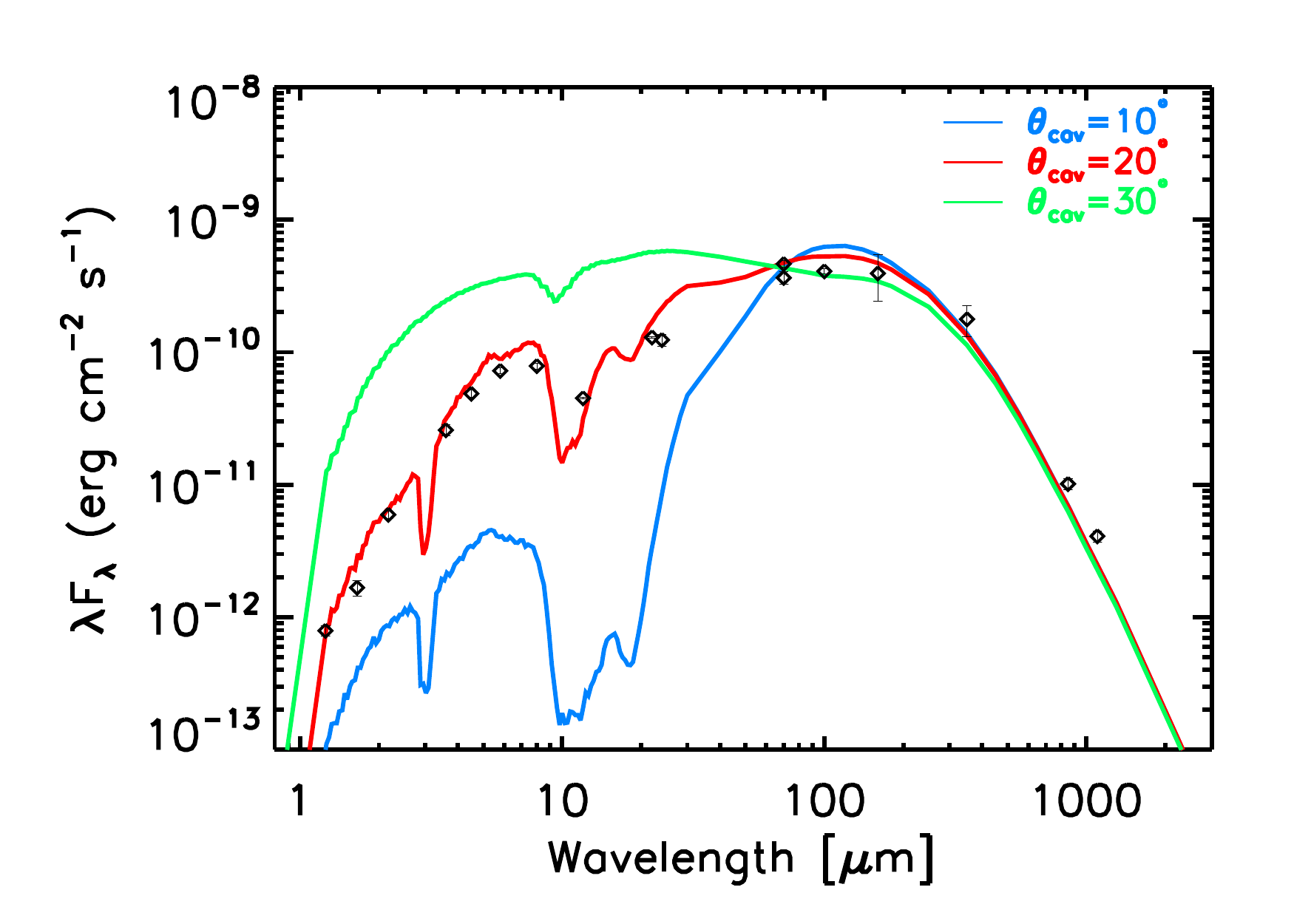}}
   \subfigure{\includegraphics[trim={0.5cm 0.5cm 0.5cm 0.5cm}, clip, width=0.98\columnwidth, keepaspectratio]{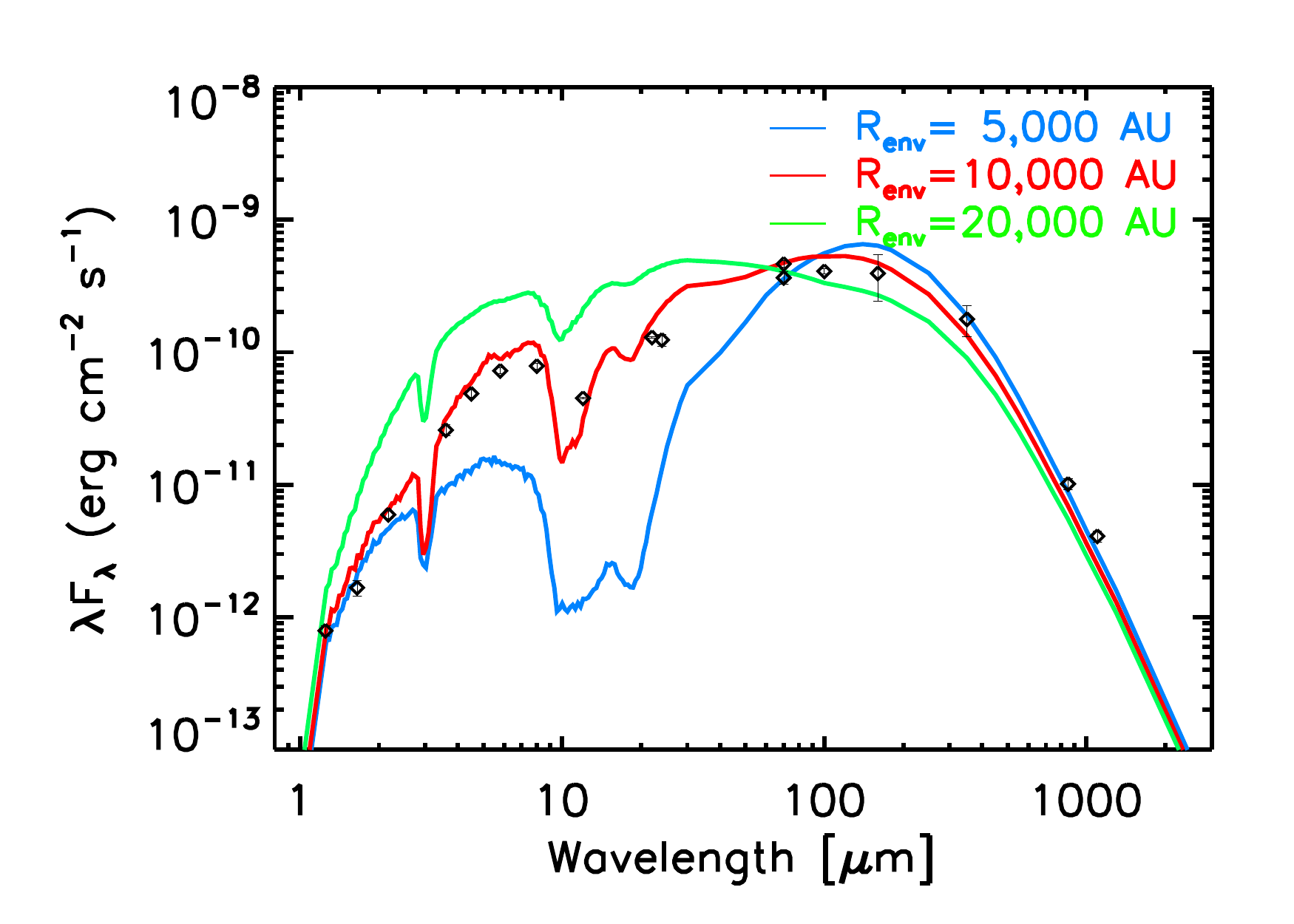}}
\caption{Parameter exploration showing the effect of each parameters. Top, middle, and bottom panels show the computed SEDs with varying power-law index, cavity opening angle and envelope radius, respectively. The black diamonds represent the observed fluxes.}
\label{fig:2d_parameter_exploration}
\end{figure}

From the fiducial model, we show the effects of three parameters on SEDs (Figure \ref{fig:2d_parameter_exploration}). 
The top panel shows the model SEDs with varying envelope density power-law index. 
As the envelope density power-law index varies, despite small differences of the SED peak flux near $\sim$100 $\mu$m, the overall SED is not affected. In order to find the power-law index of the envelope density profile, another constraint is needed (see Section \ref{sec:rip}).
However, the variation increases as the envelope size and the cavity opening angle become bigger and smaller, respectively, compared to the fiducial model (the top panel of Figure \ref{fig:2d_27cases}).

The middle panel in Figure \ref{fig:2d_parameter_exploration} shows the SEDs with varying cavity opening angles. As the opening angle becomes larger, NIR and MIR fluxes are enhanced and the flux at the SED peak is reduced. This is because more photons from the protostar and the disk can escape scattering along the cavities with a larger opening angle.

The bottom panel in Figure \ref{fig:2d_parameter_exploration} shows the SEDs with varying envelope size. As the envelope size becomes larger, the NIR and MIR fluxes increase. 
The fixed envelope mass is the cause; a bigger envelope size is regulated by a lower density to keep the envelope mass the same.
Thus the larger envelope is more transparent than the smaller one, allowing more shorter wavelength photons to escape from the system.

\begin{figure}
   \centering
   \subfigure{\includegraphics[trim={0.5cm 0.5cm 0.5cm 0.5cm}, clip, width=0.98\columnwidth, keepaspectratio]{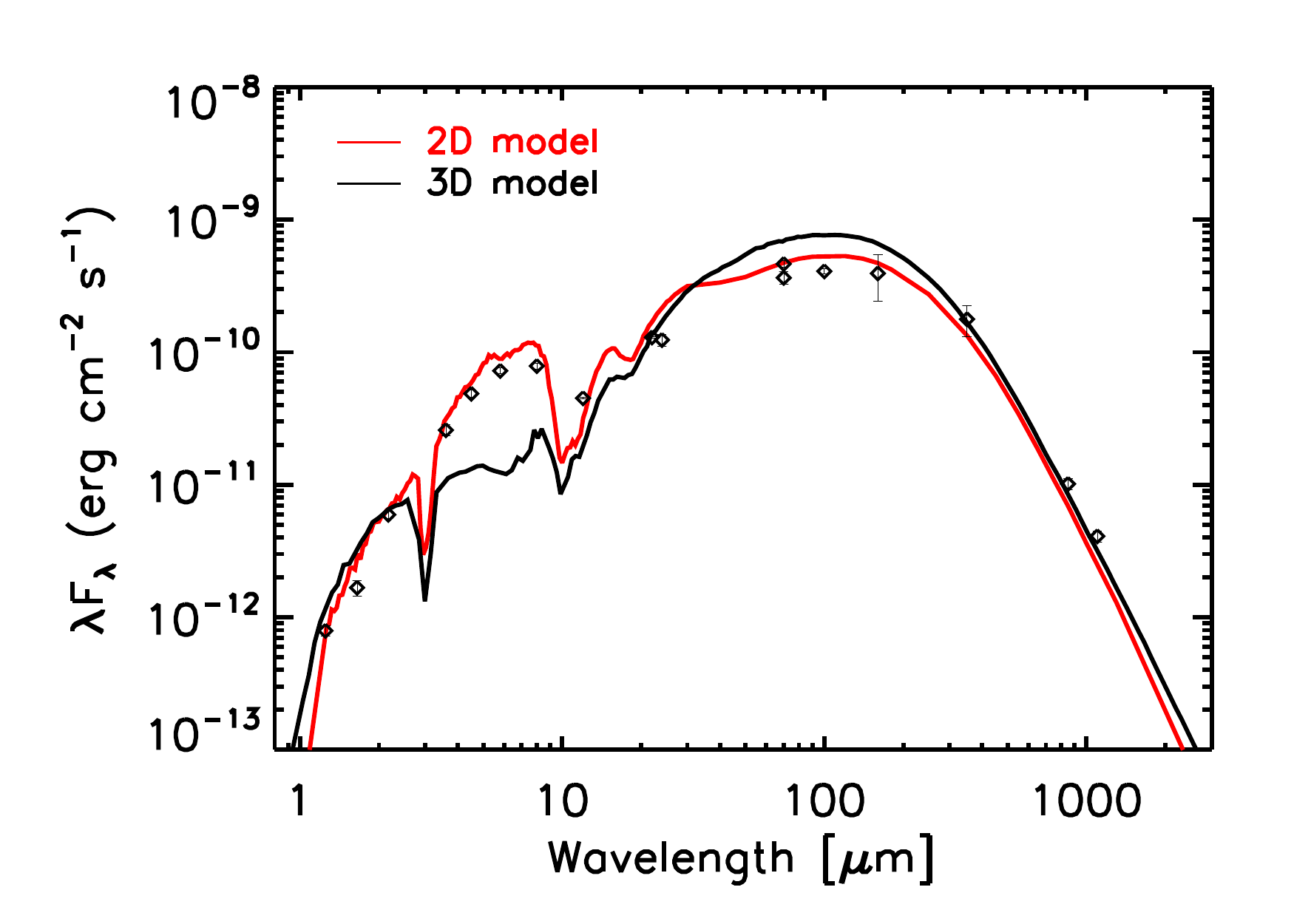}}
\caption{ SEDs for the 2D (red) and 3D (black) modelings of EC 53, adopting model parameters from the fiducial model. Solid lines represent radiative transfer calculations with the protostar as the sole luminosity source in MCRT. Black diamonds represent observed SED of EC 53.}
\label{fig:fiducial}
\end{figure}

\begin{figure}
   \centering
   {\includegraphics[trim={0.3cm 0.3cm 0.5cm 0.3cm},clip,width=85mm]{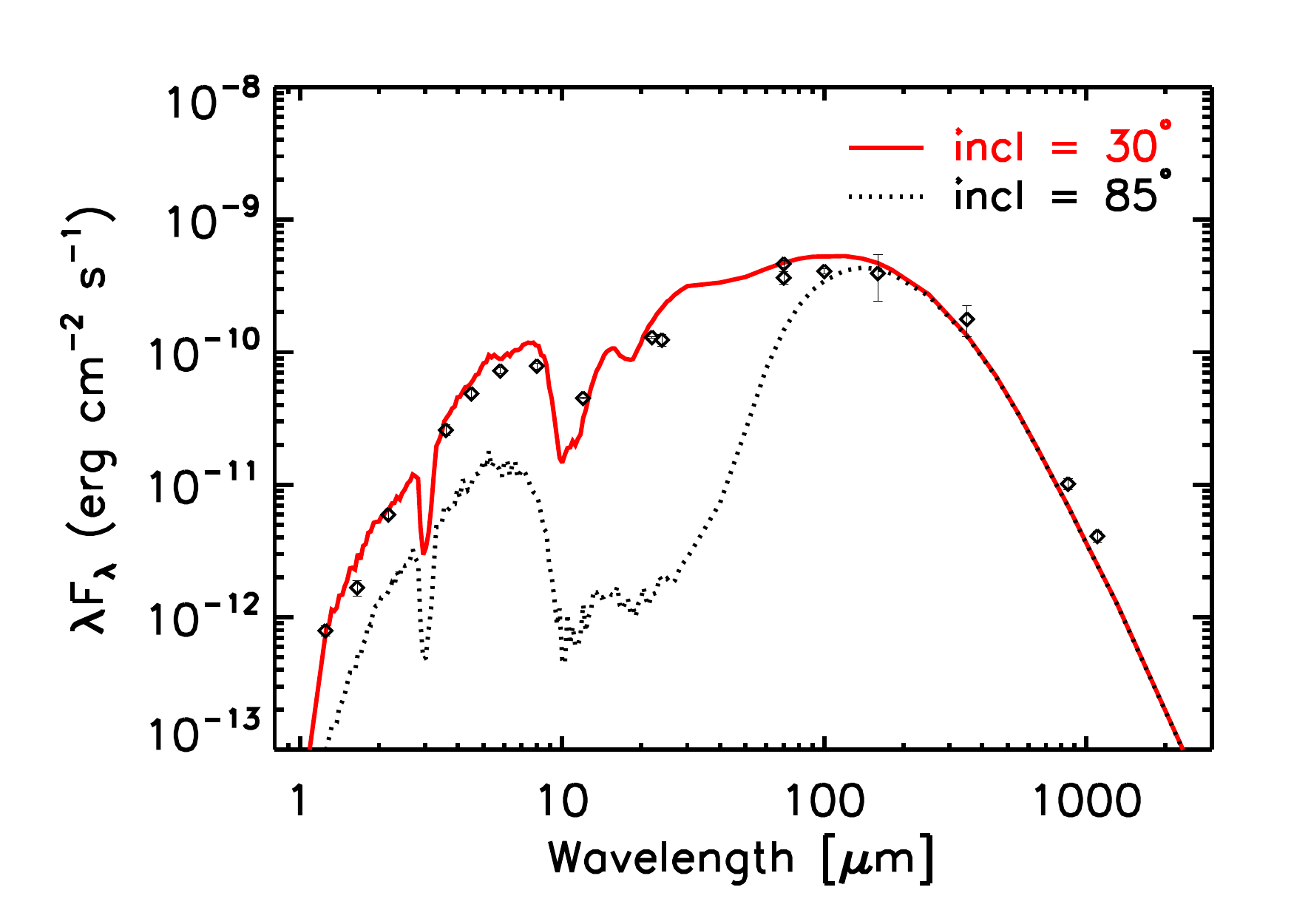}}
\caption{The 2D SEDs with different inclination angles $\theta_{inc}$. Solid line shows the model with nearly face-on (30$^{\circ}$) and dotted line shows that in the edge-on view (85$^{\circ}$).}
\label{fig:2d_incl}
\end{figure}

Our models assume that the disk inclination is nearly face-on (30$^{\circ}$), which is consistent with the inclination measured from a recent ALMA observation \citep{lee20}. The models with an inclination of $30^{\circ}$ reproduce the SED well from NIR to sub-mm wavelengths. \citet{hodapp99} and \citet{hodapp12} inferred that the system is nearly edge-on because of a cometary nebula seen in the NIR images with a dispersed shape. 
\citet{dionatos10} also reported the tentative detection of two weak (blue- and red-shifted) lobes of $^{12}$CO J = 3 $\rightarrow$ 2 pointing to the south-east and north-west from SMM5 (= EC 53).  However, in our study, the models with an edge-on disk lead to significant absorption in the IR, which is not consistent with the observations (Figure \ref{fig:2d_incl}). 

\subsection{3D model}\label{sec:3d_sed_fit}
Taking into account the results from the 2D modeling presented in  \S~\ref{sec:2d_sed_fit}, we also run the RT modeling using the 3D density profile as described in Section~\ref{sec:3d_model}. The advantage of using a 3D density profile is that the envelope asymmetries can be incorporated into the radiative transfer models (see Figure~\ref{fig:3d_col_dens}, left). While we cannot control the steepness of the radial density profile (as this is produced self-consistently in the hydrodynamic simulation), the envelope and disk parameters may be modified to match those of the 2D fiducial model. We seed a cavity with properties as described in \S~\ref{sec:2d_model} and limit the computational domain to $10,000 \ \text{AU}$, as in the fiducial model. The density is then rescaled so that the total gas mass of the YSO remains as $5.8 \ \text{M}_\odot$. 

In Figure~\ref{fig:fiducial}, we compare the SEDs computed from the 2D and 3D models with the observed SED. 
Both model SEDs match the observed fluxes at (sub-) mm. 
The SED calculated from the 3D model shows lower fluxes at NIR and higher fluxes at FIR wavelengths than those calculated from the 2D model because of differences in the envelope and disk density structures.
The envelope density from self-consistent formation of the protostar and disk is more concentrated to the center, so the disk midplane is denser in the 3D model than the 2D model.
Therefore, the 3D model is affected more by the extinction; the shorter wavelength photons from the protostar are absorbed more by the inner envelope material in the 3D model. The absorbed photons are reprocessed in the envelope and radiated at longer wavelengths. According to this comparison, the steeper envelope density profile (with the approximated power-law index of $\sim$2) in the 3D model may not be appropriate for EC 53.

\section{The outburst-phase SED of EC 53}\label{sec:burst}

\begin{figure}
   \centering
   {\includegraphics[trim={0.3cm 0.3cm 0.5cm 0.3cm},clip, width=0.98\columnwidth, keepaspectratio]{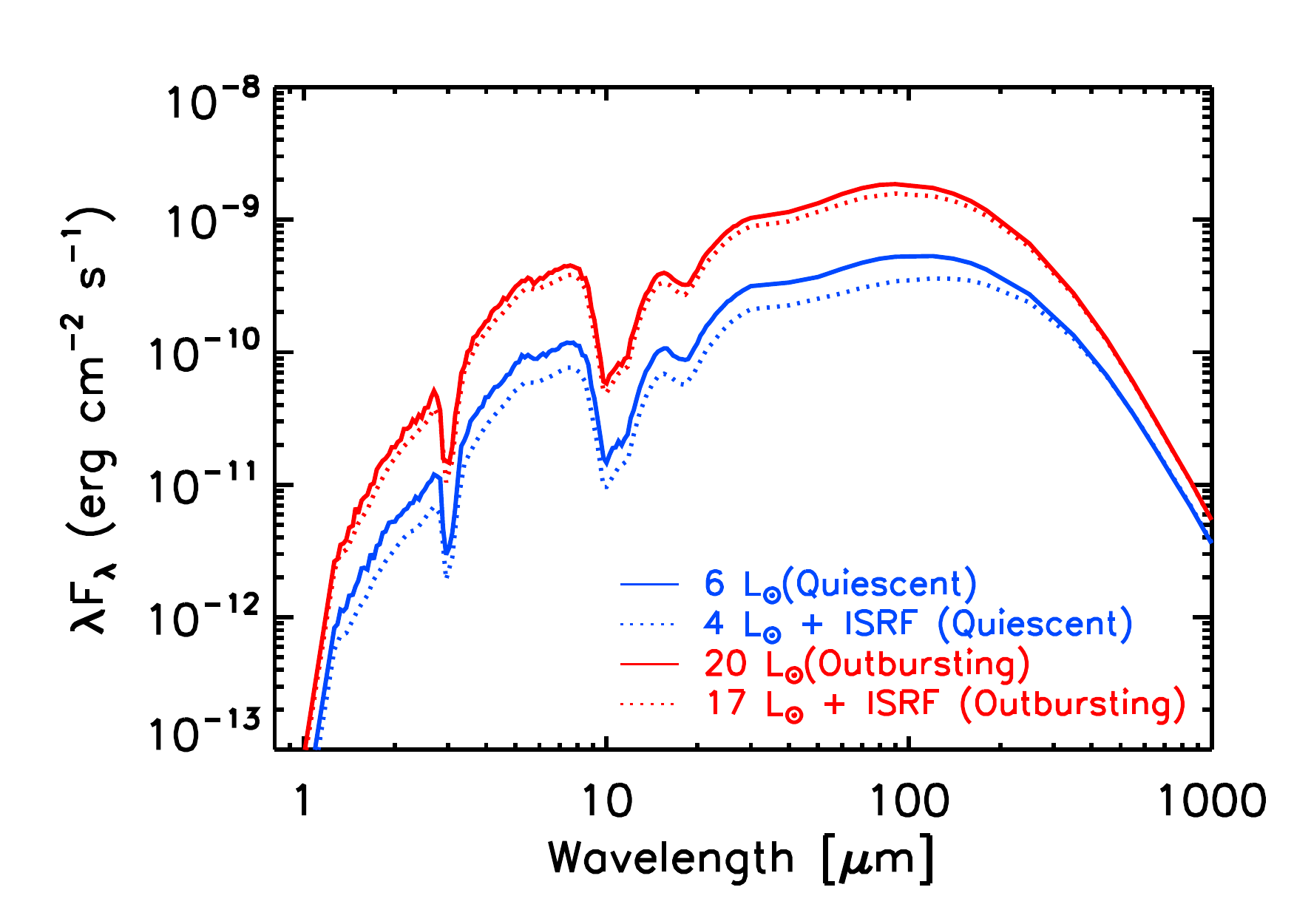}}
\caption{SEDs for modeling of EC 53, between quiescent (blue lines) and outburst (red lines) phases. Solid and dotted lines represent radiative transfer calculations to fit the observed 850 $\micron$ flux without and with the ISRF heating, respectively.}
\label{fig:sed_burst}
\end{figure}

We increase the luminosity of the central protostar to reproduce the flux increase of $\sim1.5$ at 850 $\mu\text{m}$, as observed by the JCMT Transient survey \citep{yoo17}. 
For the 2D fiducial model, when the heating from the ISRF is not taken into account, an outburst protostellar luminosity of L$_{\rm o,*} = 20 \ \text{L}_\odot$ reproduces a flux increase by a factor of $\sim1.5$ between quiescent and outbursting phase. This corresponds to an internal luminosity rise by a factor of $\sim3.3$, slightly smaller than the ratio  ($\sim3.5$) estimated by \citet{yoo17}.

\begin{figure}
   \centering
   {\includegraphics[trim={0.3cm 0.3cm 0.5cm 0.3cm},clip, width=0.98\columnwidth, keepaspectratio]{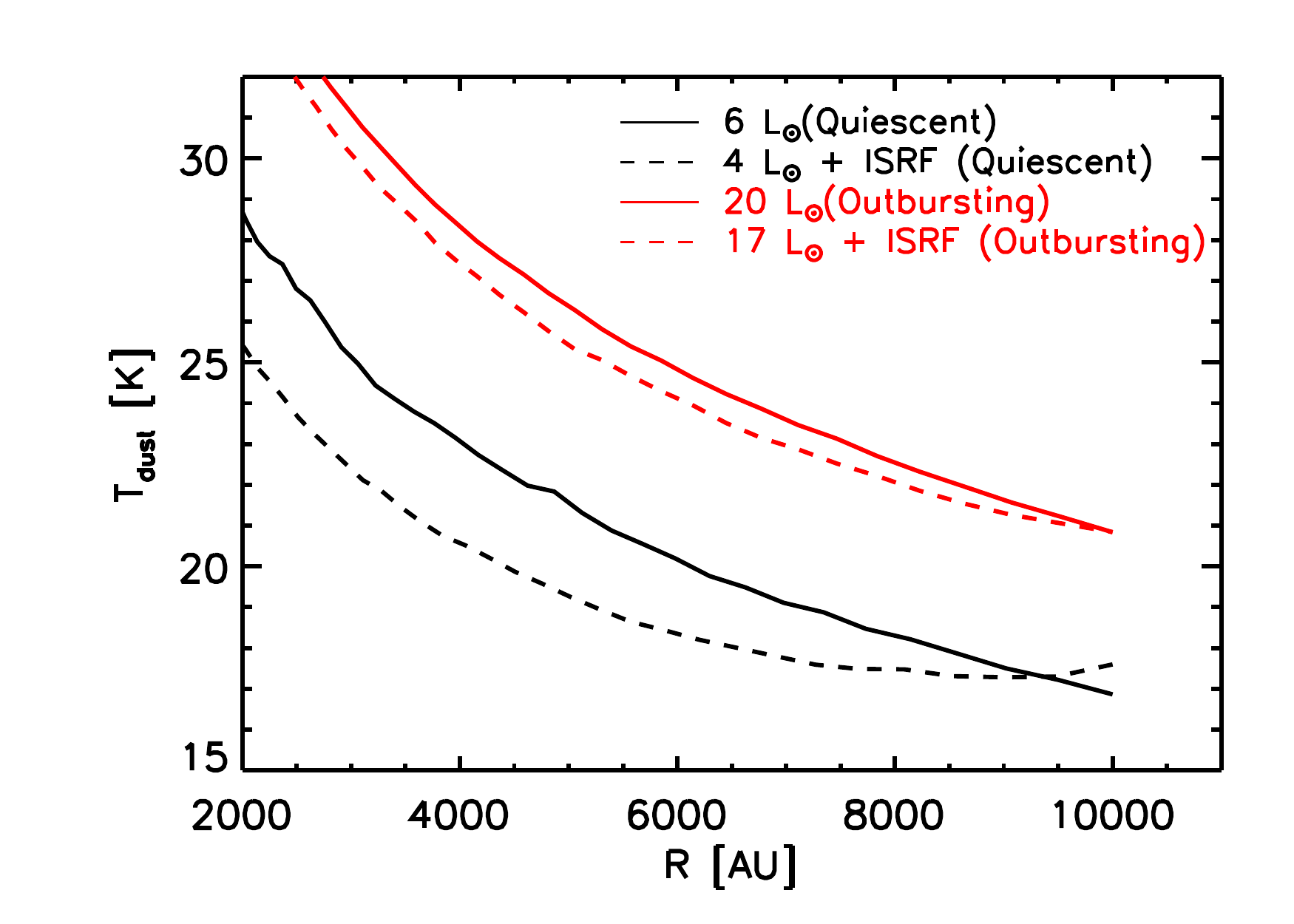}}
\caption{The radial temperature profiles of 2D modelings of EC 53 for the quiescent (black) and outburst (red) phases. The solid and dotted lines represent the temperature profiles of without and with the ISRF heating, respectively.}
\label{fig:r_vs_t}
\end{figure}

Figure~\ref{fig:sed_burst} presents SEDs for the 2D models for quiescent (blue) and outbursting (red) phases.
Dotted and solid lines represent SEDs with (dotted) and without (solid) ISRF contributions, respectively. 
When the ISRF is included, the required internal luminosities (4 and 17 $\text{L}_\odot$ for the quiescent and outbursting phases, respectively) to fit the SEDs are lower because the outer envelope can be heated by the ISRF as well as the central source. However, a greater luminosity rise (by a factor of ~4.3) is needed to increase the 850 $\micron$ flux by a factor of 1.5 since the external heating by the ISRF remains constant.  
The radial temperature distributions of 2D models in Figure \ref{fig:r_vs_t} show that the temperature at large radii does not drop much when the external heating by the ISRF is included.

To describe the effect of ISRF in a different way, in Table~\ref{tab:ratios}, we present the flux response at 450 and 850~$\mu\text{m}$ to the same change in protostellar luminosity (L$_{\rm*} = 6$ and $20 \ \text{L}_\odot$), with and without heating from the ISRF. The flux ratios are shown between outbursting and quiescent phases for the fiducial 2D and 3D models. 
When the external heating from ISRF is included, the flux ratios at 450 and 850 $\mu$m are reduced by $\sim$10 $\%$. 
\citet{macfarlane19a} explored the role of external heating by ISRF for Class 0/I protostars with a similar envelope mass (5.4 M$_{\odot}$) but a much higher luminosity enhancement (L$_{\lambda\text{,o}}$/L$_{\lambda\text{,q}}$ $>$ 550). They concluded that the ISRF attenuates the flux increase at long wavelengths if the fluxes are measured with a large aperture that includes the outer envelope, which is easily heated by the ISRF.

\begin{table}
	\caption{Flux ratios between quiescent and burst phases at $450$ and $850 \ \mu\text{m}$. The ratios are defined as the outbursting flux ($F_{\lambda\text{,o}}$) divided by quiescent flux ($F_{\lambda\text{,q}}$). For both 2D and 3D models, the protostellar luminosities for quiescent and outbursting phases are L$_{\rm*} = 6$ and $20 \ \text{L}_\odot$, respectively. Two cases with and without the external heating by the ISRF are presented.}
	\renewcommand{\arraystretch}{1.5} 
	\begin{tabular}{ c c c c}
		\hline
		Model & Luminosity & $F_{\lambda\text{,o}} / F_{\lambda\text{,q}}$ & $F_{\lambda\text{,o}} / F_{\lambda\text{,q}}$ \\
		dimensionaliry & Source(s) & (450 $\mu$m) & (850 $\mu$m) \\
		\hline
		 \multirow{2}{*}{2D} & Protostar & 1.87 & 1.55 \\
		 &Protostar + ISRF & 1.69 & 1.42 \\
		\multirow{2}{*}{3D} & Protostar & 1.81 & 1.54 \\
		 &Protostar + ISRF & 1.70 & 1.44 \\
		\hline
	\end{tabular}
\label{tab:ratios}
\end{table}

\begin{figure}
   \centering
   {\includegraphics[trim={0.3cm 0.3cm 0.5cm 0.3cm},clip, width=0.98\columnwidth, keepaspectratio]{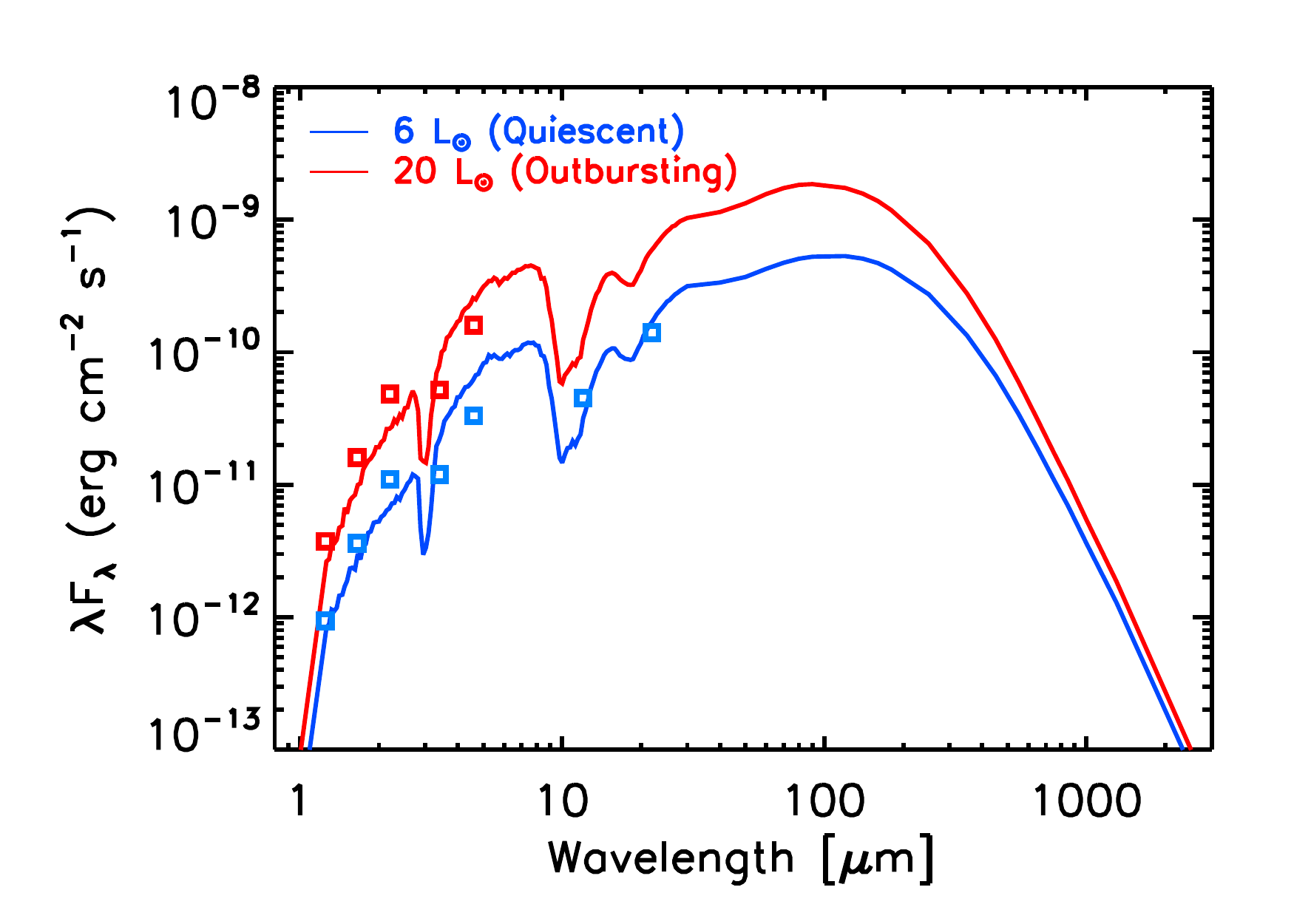}}
\caption{The 2D modeling SEDs for both in the quiescent (blue) and outburst (red) phases without the external heating by the ISRF. UKIRT J, H and K bands and WISE 3.4 and 4.6 $\mu$m photometric data are presented both in the quiescent and outburst phases. WISE 12 and 22 $\mu$m data are presented only in quiescent phase. Photometric uncertainties for all data points are less than 3$\%$ of their fluxes.}
\label{fig:nir}
\end{figure}

\begin{table}
	\caption{Observed flux ratios between quiescent and burst phases of EC 53. The ratios are defined as the outbursting flux ($F_{\lambda\text{,o}}$) divided by quiescent flux ($F_{\lambda\text{,q}}$).}
\centering
	\begin{tabular}{ c c c}
		\hline
		Instrument &  Wavelength [$\mu$m] & $F_{\lambda\text{,o}} / F_{\lambda\text{,q}}$ \\
		\hline
		 UKIRT/J & 1.25 &3.99 \\
		 UKIRT/H &1.65  & 4.42\\
		 UKIRT/K & 2.2 & 4.41\\
		 WISE/W1 & 3.4 & 4.34 \\
		 WISE/W2 & 4.6 & 4.81\\
		\hline
	\end{tabular}
\label{tab:ratios_obs}
\end{table}

Figure \ref{fig:nir} presents the best-fit 2D SEDs for quiescent (blue) and outbursting (red) phases with the NIR and MIR photometric observations. The NIR photometric data were obtained at J, H, and K bands with the United Kingdom Infra-Red Telescope (UKIRT) taken on 30 April, 2016 and 10 October, 2016 for the quiescent and outburst phases, respectively (Y.-H. Lee et al., in prep.). The MIR data were collected from WISE and NEOWISE surveys at 3.4, 4.6, 12 and 22 $\mu$m taken on 27 March, 2016 for the quiescent phase and at 3.4 and 4.6 $\mu$m taken on 28 March, 2017 for the outburst phase (Contreras Pe\~{n}a et al., submitted; Table \ref{tab:obs}). Since EC 53 shows flux variations with the $\sim$543 days periodicity, the NIR and MIR data are selected to represent the quiescent and outburst phase fluxes within the same phase of the 850 $\mu$m study of \citet{yoo17}. The flux ratios between quiescent and burst phases between the two phases are listed in Table \ref{tab:ratios_obs}. 
The flux enhancement factors at NIR and MIR wavelengths are much larger than that at sub-mm wavelengths. The NIR/MIR flux variation follow more closely changes in the source luminosity while the submm flux variation more closely traces the temperature variation of dust grains in the envelope (\citealt{johnstone13}, Contreras Pe\~{n}a et al., submitted).
The NIR and MIR periodic variations with monitoring observations will be presented in detail in a separate study (Y.-H. Lee et al., in prep.) 

Our best-fit model for the burst phase is reasonably consistent with the observed NIR and MIR photometry, although the model SED was constrained by the flux enhancement and the radial intensity profile only at 850 $\mu$m. The small difference between observations and the model at NIR and MIR could be adjusted by modifying the disk and cavity properties (Contreras Pe\~{n}a et al., submitted), which can be done in a future study. 

Our models display a narrow absorption dip at 3.1 $\mu$m due to water ice and broad absorption dips at 10 $\mu$m due to silicates. Although neither are constrained by current observations, future NIR and MIR spectroscopic observations could use these absorption bands to test the dust properties in the envelope/disk system as well as their physical characteristics affected by the episodic accretion process more precisely. 

\section{RADIAL INTENSITY PROFILE}\label{sec:rip}
\begin{figure*}
   \centering
   \subfigure{\includegraphics[trim={0.5cm 1cm 1cm 1cm}, clip, width=0.98\columnwidth, keepaspectratio]{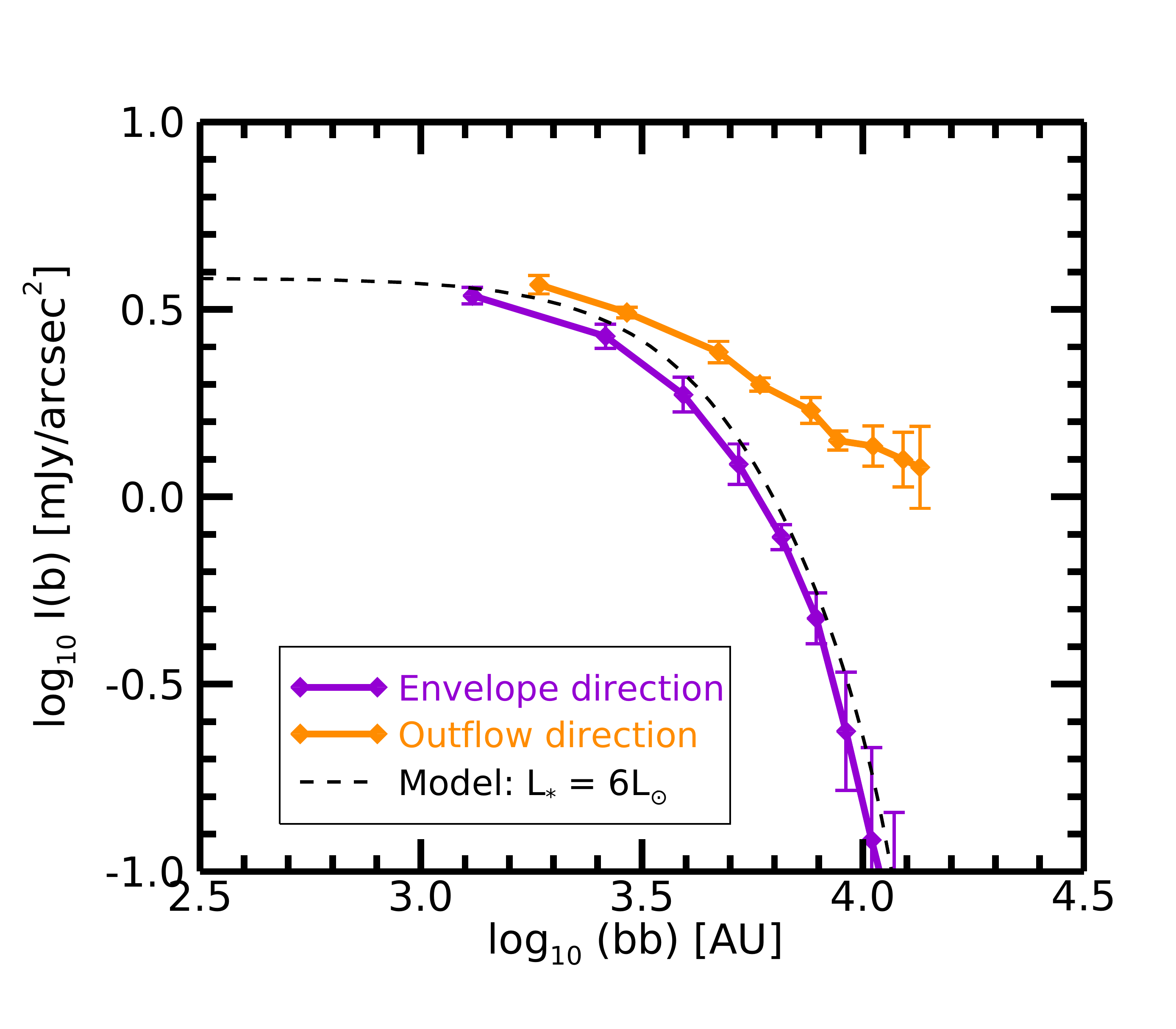}}
   \subfigure{\includegraphics[trim={0.5cm 1cm 1cm 1cm}, clip, width=0.98\columnwidth, keepaspectratio]{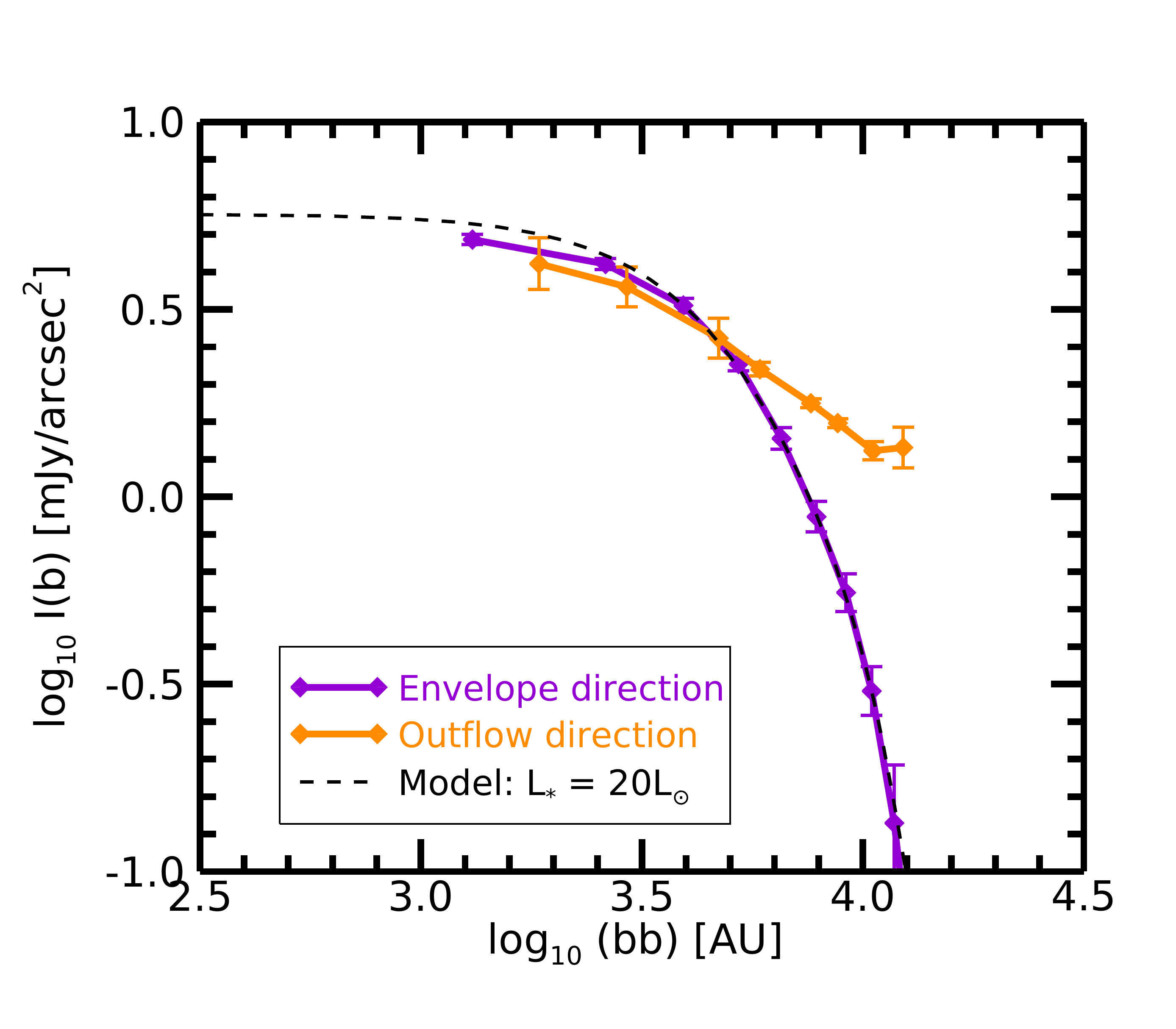}}
\caption{The 850 $\mu$m radial intensity profiles of EC 53 for quiescent (left) and outbursting (right) phases. Purple and orange solid lines present the radial intensity profiles along the envelope (P.A.= 90$^{\circ}$) and outflow cavity (P.A.= 330$^{\circ}$) directions, respectively. The dashed line is the modeled radial intensity profiles (2D) without ISRF from our fiducial model.}
\label{fig:rad_direction}
\end{figure*}

The radial intensity profile has been used to study the detailed structure of envelopes of embedded YSOs (e.g., \citealp{chandler00,shirley02}). 
The envelope density and temperature profiles could be directly investigated using the sub-mm radial intensity profiles since the envelope becomes optically thin at sub-mm wavelength. 
Our SED analysis shows that various parameters of the envelope density structure are constrained reasonably well with the observed SED. However, the power-law index of the envelope density profile is not well constrained; the modeling results with three power-law indices are not significantly different in fiducial 2D models for a given cavity opening angle (see Section \ref{sec:fid} and the top panel of Figure \ref{fig:2d_parameter_exploration}).

In this section, we investigate the radial intensity profile of the SCUBA-2 850 $\mu$m image. 
For comparison, the radiative transfer models with and without the external heating by the ISRF are calculated, for both the quiescent and outbursting phases.  The envelope density structures are described by simple power-law functions. 

\subsection{Observed radial intensity profile}

In this subsection, we examine the radial intensity profiles along directions of outflow cavity and dense envelope in images of EC 53. We use the JCMT/SCUBA-2 850 $\mu$m continuum images observed on 2 February, 2016 (quiescent phase) and 22 February, 2017 (outburst phase).
The disk direction of EC 53 is adopted from the analysis of a high-resolution ALMA observation (P.A.= 60$^{\circ}$, where the position angle (P.A.) is measured relative to the north pole;  \citealt{lee20}). 
The outflow cavity direction (P.A.= 330$^{\circ}$) is assumed to be perpendicular to the disk direction. 
However, we extract the radial intensity profile of the envelope along P.A.= 90$^{\circ}$ to include more pixels in the images.
Along each direction, the intensity at a given radial distance from the central star is obtained by averaging the intensities over adjacent three pixels. 
The uncertainty of intensity is calculated by the standard deviation of the intensities from three pixels because the calibration and measurement errors are much smaller than the intensity variation at different pixels.

Figure \ref{fig:rad_direction} compares the radial intensity profiles from observations with the 2D models ($\rho \propto$ r$^{-1.5}$) in both quiescent (left) and outbursting (right) phases. 
Only the radial intensity profile measured along the envelope direction shows clear change; the intensity profile along the envelope direction is enhanced in the outburst phase while that along the outflow cavity direction does not show a notable variation.

The enhanced 850 $\mu$m intensity profile only along the envelope direction strongly supports that the brightness increase observed in EC 53 is caused by the heated envelope through the accretion burst.
If an accretion burst occurs in an embedded YSO like EC 53, the enhanced radiation at short wavelengths is absorbed by the surrounding dense material. 
The absorbed radiation increases the temperature of the surrounding disk and envelope until the radiation escapes from the system  at longer wavelengths \citep{johnstone13}.
However, if the sub-mm flux enhancement is caused by an external source such as variable nearby bright stars,
radial intensities both along the envelope and cavity directions should be enhanced, which is not the case of EC 53.

An interesting aspect is that, in the quiescent phase, the intensity profile along the outflow cavity direction is higher than that along the envelope direction, which is not expected at 850 $\mu$m. 
Moreover, the intensity profile does not drop sharply at the boundary, but it has a much shallower profile compared to that along the envelope direction.
These features strongly suggest that the outflow cavities may be contaminated by emission from external sources.
EC 53 is located in the north-eastern periphery of the filamentary structure of the Serpens main cloud. Thus the outflow cavity region could coincide with a filamentary structure of the cloud in a relatively higher density.  
In addition, the intensity in the western direction could be affected by the nearby bright sources, which are located at the west side from EC 53 (Source 1 and 3 in Figure 1 in \citealt{yoo17}). 
Therefore, to avoid any external effects, the radial intensity profiles only along the envelope direction are considered in further analysis.
We note that although the intensity inside the cavity is contaminated by the external sources, 
the flux integrated over the cavity is not large enough to affect the total flux for the SED analysis.

\begin{figure}
   \centering
   {\includegraphics[trim={0.3cm 0.3cm 0.5cm 0.3cm},clip, width=0.98\columnwidth, keepaspectratio]{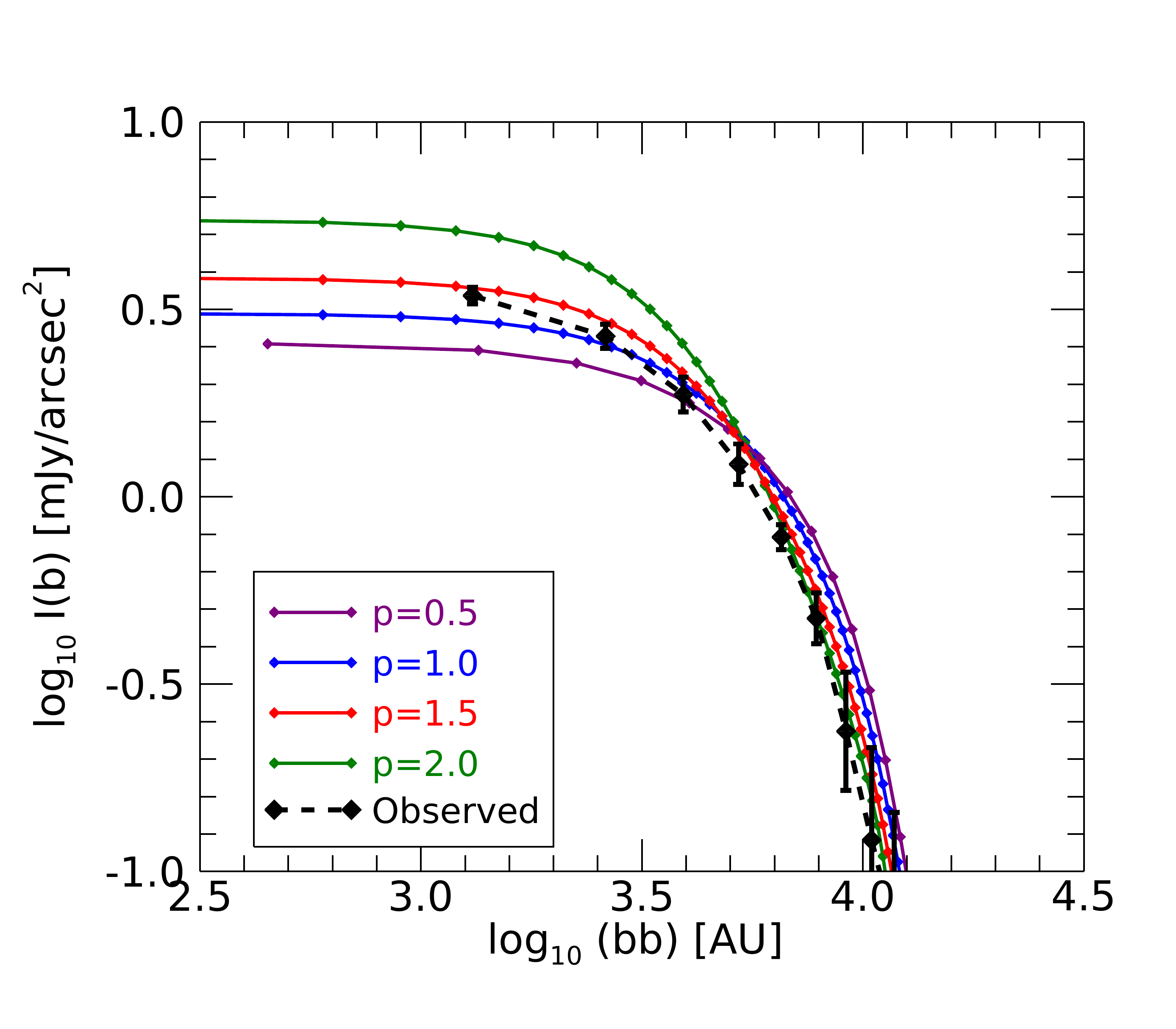}}
\caption{Modeled radial intensity profiles (2D) with varying density power-law slopes of the envelope. Purple, blue, red, and green solid lines represent the models with envelope power-law indices p of 0.5, 1.0, 1.5 and 2.0 in Equation \ref{eq:envelope}, respectively. 
Black dashed line represents the observed radial intensity profile.}
\label{fig:pw}
\end{figure}

\subsection{Modeled radial intensity profile}

The synthetic 850 $\mu$m images of EC 53 are produced using the parameters from the SED models. 
The radial intensity profiles from the synthetic images are obtained with the same method as used in the observed images.
Figure \ref{fig:pw} displays the radial intensity profile during a quiescent phase for envelope density structures that follow different power-laws.  As the density structure become shallower, more material is located at the outer envelope. As a result, the intensity becomes weaker at smaller radii and stronger at larger radii.
The best-fit model {\bf (with the $\chi^{2}_{red}$ value of 1.14)} has the envelope density distribution of $\rho \propto$ r$^{-1.5}$. 

The derived envelope density distribution of $\rho \propto$ r$^{-1.5}$ is consistent with that of the infalling envelope \citep{shu77,terebey84}.
Similarly, \citet{chandler00} and \citet{shirley02} used the JCMT/SCUBA sub-mm images to fit the radial intensity profiles of many Class 0/I YSOs with the power-law density profiles, concluding that the power-law indices of envelope structure of Class I protostars are p=1.5--2 ($\rho \propto$ r$^{-p}$). 

\begin{figure}
   \centering
      \subfigure{\includegraphics[trim={0.5cm 1cm 1cm 1cm}, clip, width=0.98\columnwidth, keepaspectratio]{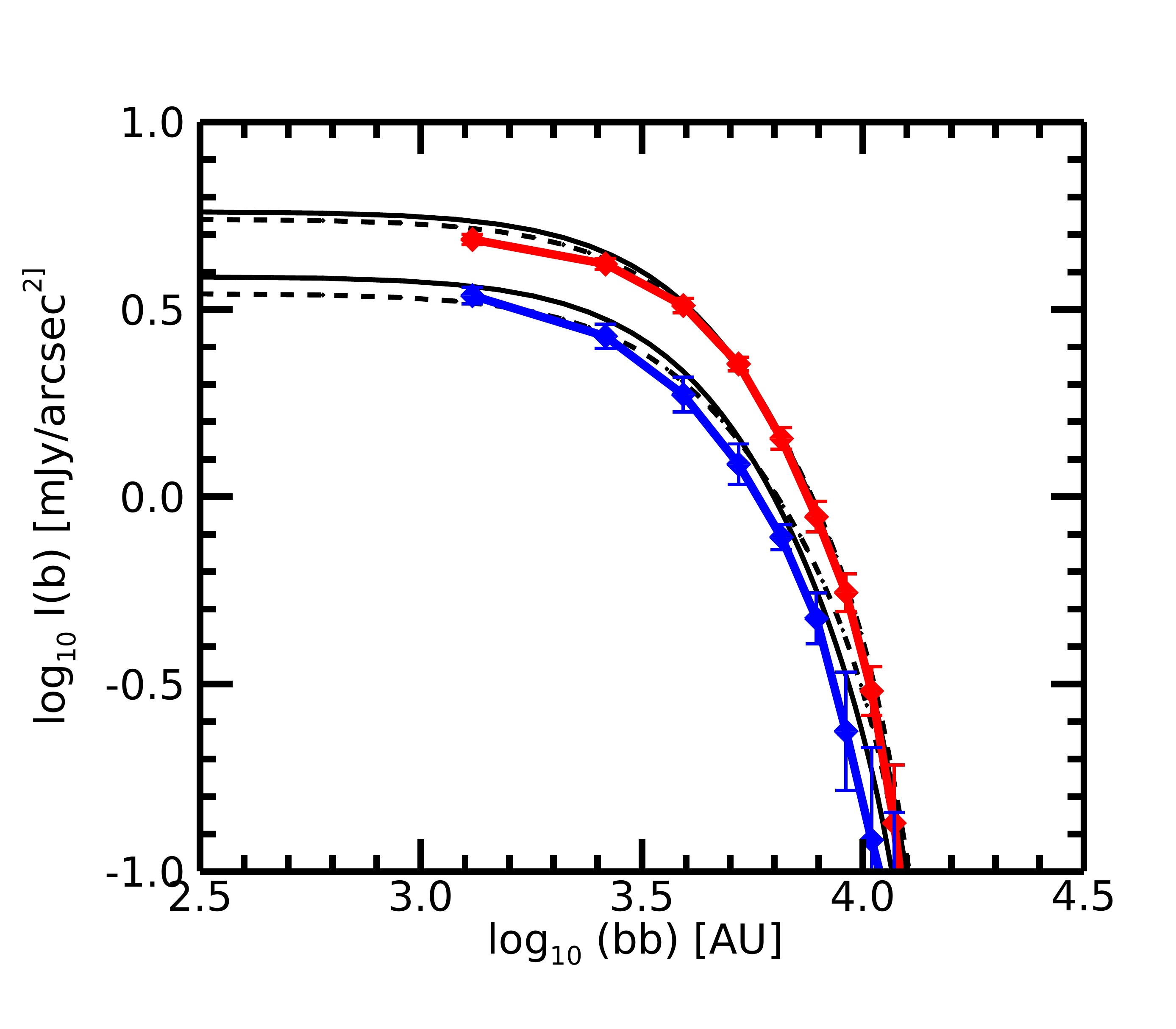}}
   \subfigure{\includegraphics[trim={0.5cm 1cm 1cm 1cm}, clip, width=0.98\columnwidth, keepaspectratio]{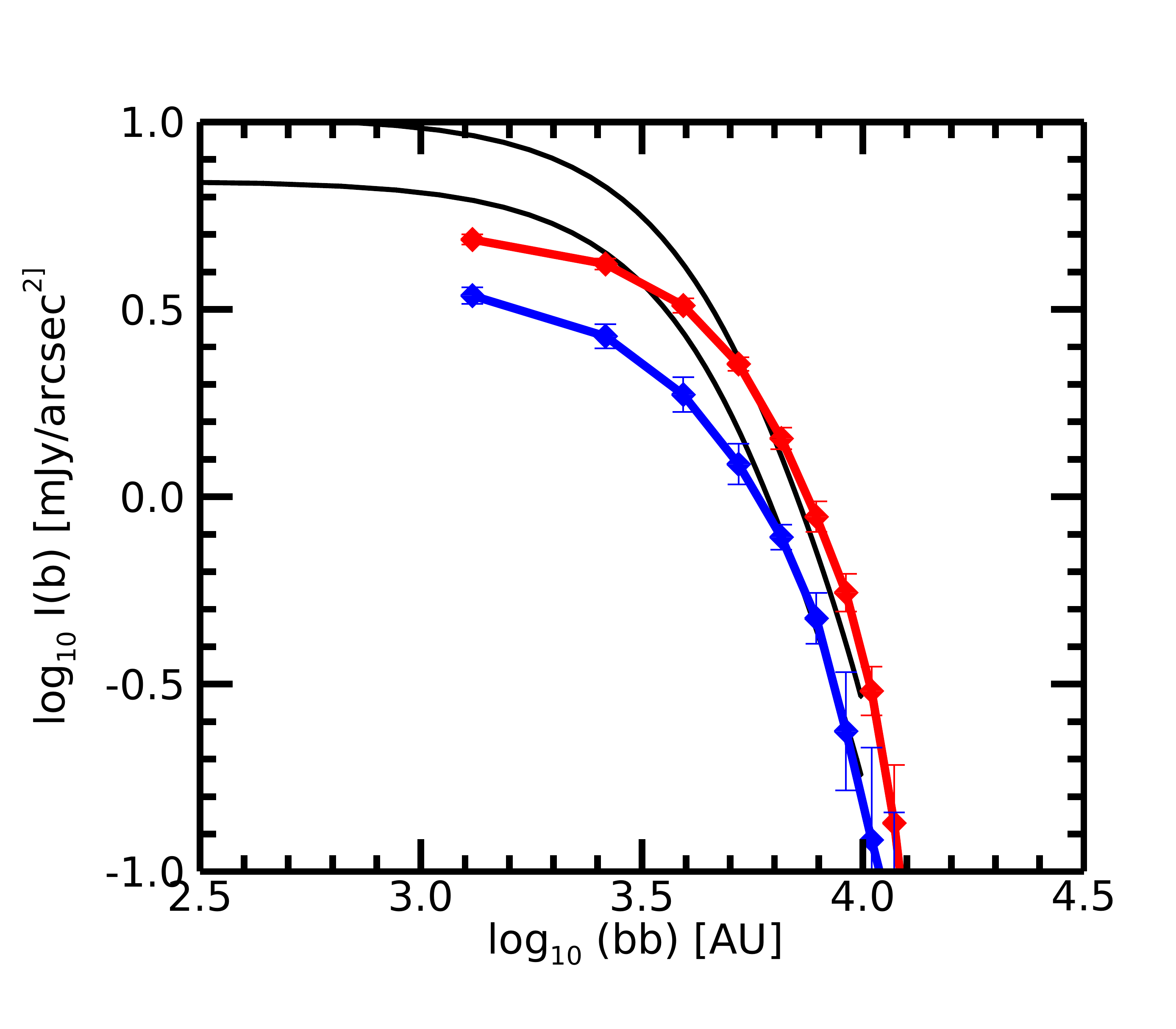}}   
\caption{ 2D (top) and 3D (bottom) radial intensity profiles for quiescent and outbursting phases of EC 53. Blue and red solid lines represent the observation for quiescent and outbursting phases, respectively. Black solid line represents the model radial intensity profile without ISRF. The black dashed lines in the top panel are the modeled radial intensity profiles with ISRF.}
\label{fig:img_fiducial}
\end{figure}

\subsection{Outburst-phase}
During the outburst, properties such as the envelope density structure, dust opacity, and characteristics of the external heating should not change considerably. Meanwhile, the enhanced internal luminosity heats up the surrounding material, increasing the temperature of the envelope. Therefore, in the outburst phase, the intensity variation of the envelope at sub-mm wavelengths is sensitive only to temperature variations \citep{johnstone13}. 

Figure \ref{fig:img_fiducial} shows the 2D (top) and 3D (bottom) radial intensity profiles in the quiescent (blue) and outbursting (red) phases, respectively.
The radial intensity profiles with (dashed lines) and without (solid lines) the ISRF are compared in the 2D model (the top panel of Figure \ref{fig:img_fiducial}), with the same parameters as in the SED models in Figure \ref{fig:sed_burst}. 
The contribution of the ISRF is insignificant for EC 53, perhaps because the envelope of EC 53 is dense enough to shield the protostar from the external UV radiation.
However, if an outbursting YSO is located in a region where the external heating is important (e.g., near a massive star), the models must include the external heating source to correctly estimate the rise of the internal luminosity. The effect of external heating from the ISRF on outbursting YSOs is discussed in \citet{macfarlane19a}. 

The beam size of SCUBA-2 ($\sim$15$\arcsec$) corresponds to the radius of $\sim$3300 AU at a distance of 436 pc. 
While the 2D models fit the radial intensity profiles very well for both the quiescent and outburst phases, the 3D models fit them only beyond $\sim$6300 AU and produce higher intensities at inner radii due to the highly concentrated density structure adopted in the 3D model.

\section{Conclusions}\label{sec:conclusions}
To understand the accretion process in star formation, it is important to study the episodic accretion in the early stages of YSOs with their thick envelopes. 
The JCMT Transient survey is measuring variability of protostars at sub-mm wavelengths, but interpreting these brightness changes in terms of source luminosities requires envelope models.
In this study, the SED and radial intensity profiles of EC 53 were modeled to quantify the observed luminosity enhancement at 850 $\mu$m, which is caused by the accretion burst.
A fiducial model to fit the SED in the quiescent phase is obtained by exploring envelope properties such as cavity opening angle, envelope size, and envelope density profile.
To reproduce the observed sub-mm brightness rise during the outburst phase, the internal luminosity should increase at least 3.3 times from that in the quiescent phase.
If the external heating is included in the model, the factor of internal luminosity rise should increase more (4.3 times) to explain the 850 $\mu$m flux enhancement.
In addition, the radial intensity profile of the SCUBA-2 850 $\mu$m image is obtained and compared with models.
Only the radial intensity profile obtained along the dense envelope direction, which does not include outflow cavities, shows significant variation, indicating that the envelope of EC 53 is heated by the accretion burst; 
the cavities, which are affected by the external radiation as well as the internal heating source, do not show much change in the radial intensity profile between the quiescent and burst phases.
According to our study, both the SED and radial intensity profile must be considered to constrain the physical properties of envelope and the enhancement factor of internal luminosity.

\acknowledgments
We thank the anonymous referee for comments that improved the paper. 
We also thank Hauyu Baobab Liu for discussions. This work was motivated and performed by the JCMT-Transient Team as part of a comprehensive effort to obtain and interpret time-domain sub-mm observations of star-forming regions. The authors appreciate the important contribution of the JCMT Transient Team members in observing, calibrating, and making available the submillimetre observations used in this paper. 
This work was supported by the Basic Science Research Program through the National Research Foundation of Korea (grant No. NRF-2018R1A2B6003423) and the Korea Astronomy and Space Science Institute under the R\&D program supervised by the Ministry of Science, ICT and Future Planning.
G. Baek was supported by NRF (NRF-2017H1A2A1043046-Global Ph.D. Fellowship Program).
BM is supported by STFC grant ST/N504014/1. DS is partly supported by STFC grant ST/M000877/1.
D.J. is supported by the National Research Council of Canada and by an NSERC Discovery Grant. 
GH is supported by grant 11773002 by the National Science Foundation of China.  
Column density images produced using the visualisation software \small{SPLASH} \citep{price07}.
Hydrodynamic simulations were performed using the UCLAN HPC facility and the COSMOS Shared Memory system at DAMTP, University of Cambridge operated on behalf of the STFC DiRAC HPC Facility. 
This equipment is funded by BIS National E-infrastructure capital grant ST/J005673/1 and STFC grants ST/H008586/1, ST/K00333X/1. 
When the data reported here were acquired, UKIRT was supported by NASA and operated under an agreement among the University of Hawaii, the University of Arizona, and Lockheed Martin Advanced Technology Center; operations were enabled through the cooperation of the East Asian Observatory.
The James Clerk Maxwell Telescope is operated by the East Asian Observatory on behalf of The National Astronomical Observatory of Japan; Academia Sinica Institute of Astronomy and Astrophysics; the Korea Astronomy and Space Science Institute; Center for Astronomical Mega-Science (as well as the National Key R\&D Program of China with No. 2017YFA0402700). Additional funding support is provided by the Science and Technology Facilities Council of the United Kingdom and participating universities in the United Kingdom and Canada. 
Additional funds for the construction of SCUBA-2 were provided by the Canada Foundation for Innovation.
\vspace{5mm}

\software{RADMC-3D \citep{dullemond12}}
\software{SPLASH \citep{price07}}

\bibliographystyle{aasjournal}
\bibliography{ms.bib} 

\end{document}